\documentclass[a4paper,11pt,fleqn]{article}
\usepackage[pdftex,colorlinks=true,linkcolor=blue,citecolor=blue,urlcolor=blue]{hyperref}

\usepackage[T1]{fontenc}
\usepackage[utf8]{inputenc}
\usepackage[english]{babel}

\usepackage[a4paper,top=17mm,bottom=28mm,inner=25mm,outer=25mm]{geometry}

\usepackage{fourier}
\usepackage{setspace} 

\usepackage{graphicx,xcolor}
\graphicspath{{images/}}

\usepackage{subfig}
\usepackage{microtype}

\usepackage{xr}
\usepackage{color}
\usepackage[explicit]{titlesec}
\usepackage{multicol}
\usepackage{amsmath}
\usepackage{amssymb}
\usepackage{bm}
\usepackage{float}
\usepackage{placeins}
\usepackage{relsize}
\usepackage{wrapfig}
\usepackage[numbers,sort&compress,square]{natbib}
\usepackage{mathtools}
\usepackage[version=3]{mhchem}

\usepackage{graphics}
\usepackage{braket}

\usepackage{enumitem}
\setlist[itemize]{itemsep=-3pt, topsep=3pt, leftmargin=2.5em}
\setlist[enumerate]{itemsep=-3pt, topsep=3pt, leftmargin=2.5em}
\setenumerate[1]{label={(\arabic*)}}

\newcommand*{\myeqref}[2][Eq.~]{#1(\ref{#2})}

\newcommand*{\myfigref}[2][Fig.~]{#1\ref{#2}}

\newcommand*{\mysecref}[2][Sec.~]{#1\ref{#2}}

\newcommand{\todo}[1]{\textcolor{blue}{TODO: #1}}

\newcommand*{\appfigref}[2][Fig.~]{#1\ref{#2} of the SI~\cite{kahle_supplementary}}
\newcommand{\appsecref}[2][Sec.~]{#1\ref{#2}}
\newcommand{\suppltabref}[1]{Table~\ref{#1} of the SI~\cite{kahle_supplementary}}

\newcommand{\supplref}[2][Fig.~]{#1\ref{#2} of the SI~\cite{kahle_supplementary}}

\newcommand*{\articletitle}[3]{
  \begin{center}
    {\large \bfseries #1}\\[10pt]
    {#2}\\[5pt]
    {}
  \end{center}
}

\newenvironment{myabstract}{
    \begin{center}
    \begin{minipage}[]{0.9\hsize}
    \setstretch{1.1} 
    \hrulefill

        \textbf{Abstract:}
    }
{

	\hrulefill
	\end{minipage}
	\end{center}

	\setstretch{1.2} 
}

\begin{document}
\articletitle{High-throughput computational screening for solid-state Li-ion conductors}
{Leonid Kahle,$^a$ Aris Marcolongo,$^{a\dag}$ and Nicola Marzari$^a$}
{Theory and Simulation of Materials (THEOS), and National Centre for Computational Design and Discovery of Novel Materials (MARVEL), \'{E}cole Polytechnique F\'{e}d\'{e}rale de Lausanne, CH-1015 Lausanne, Switzerland}
\footnotetext{$^a$~Theory and Simulation of Materials (THEOS), and National Centre for Computational Design and Discovery of Novel Materials (MARVEL), \'{E}cole Polytechnique F\'{e}d\'{e}rale de Lausanne, 1015 Lausanne, Switzerland; \\ E-mail: \href{mailto:leonid.kahle@epfl.ch}{leonid.kahle@epfl.ch}}
\footnotetext{$\dag$~Present address: IBM Research--Zurich, CH-8803 R\"uschlikon, Switzerland}

\begin{myabstract}
We present a computational screening of experimental structural repositories for fast Li-ion conductors, with the goal of finding new candidate materials for application as solid-state electrolytes in next-generation batteries.
We start from $\sim$1400 unique Li-containing materials, of which $\sim$900 are insulators at the level of density-functional theory.
For those, we calculate the diffusion coefficient in a highly automated fashion, using extensive molecular dynamics simulations on a potential energy surface (the recently published pinball model) fitted on first-principles forces.
The $\sim$130 most promising candidates are studied with full first-principles molecular dynamics, first at high temperature and then more extensively for the 78 most promising candidates.
The results of the first-principles simulations of the candidate solid-state electrolytes found are discussed in detail.
\end{myabstract}

\section{Introduction}

Application of inorganic solid-state  lithium-ionic conductors as electrolytes could mitigate or overcome the severe safety challenges imposed by the use of volatile and flammable liquid or polymer electrolytes in today's Li-ion batteries~\cite{balakrishnan_safety_2006,goodenough_challenges_2010,schaefer_electrolytes_2012,janek_solid_2016,kato_high-power_2016,kwade_current_2018}.
Complete replacement of the liquid electrolyte by a solid ceramic would result in an all-solid-state Li-ion battery, highly beneficial due to the higher electrochemical stability of inorganic electrolytes, compared to their organic counterparts~\cite{kim_review_2015}.
Several structural families of promising solid-state Li-ion conductors have been researched intensely over the last decades~\cite{alpen_ionic_1977,robertson_review_1997,quartarone_electrolytes_2011,bachman_inorganic_2016}, but the many necessary criteria for successful deployment, such as fast-ionic/superionic~\cite{boyce_superionic_1979,knauth_solid-state_2002,hull_superionics:_2004} diffusion of Li ions,
very low electronic mobility, wide electrochemical stability windows, and high mechanical stability~\cite{weppner_engineering_2003,tikekar_design_2016,manthiram_lithium_2017,nanda_frontiers_2018} motivate the search for novel candidates.

\begin{wrapfigure}{R}{0.5\hsize}
    \centering
    \includegraphics[width=\hsize]{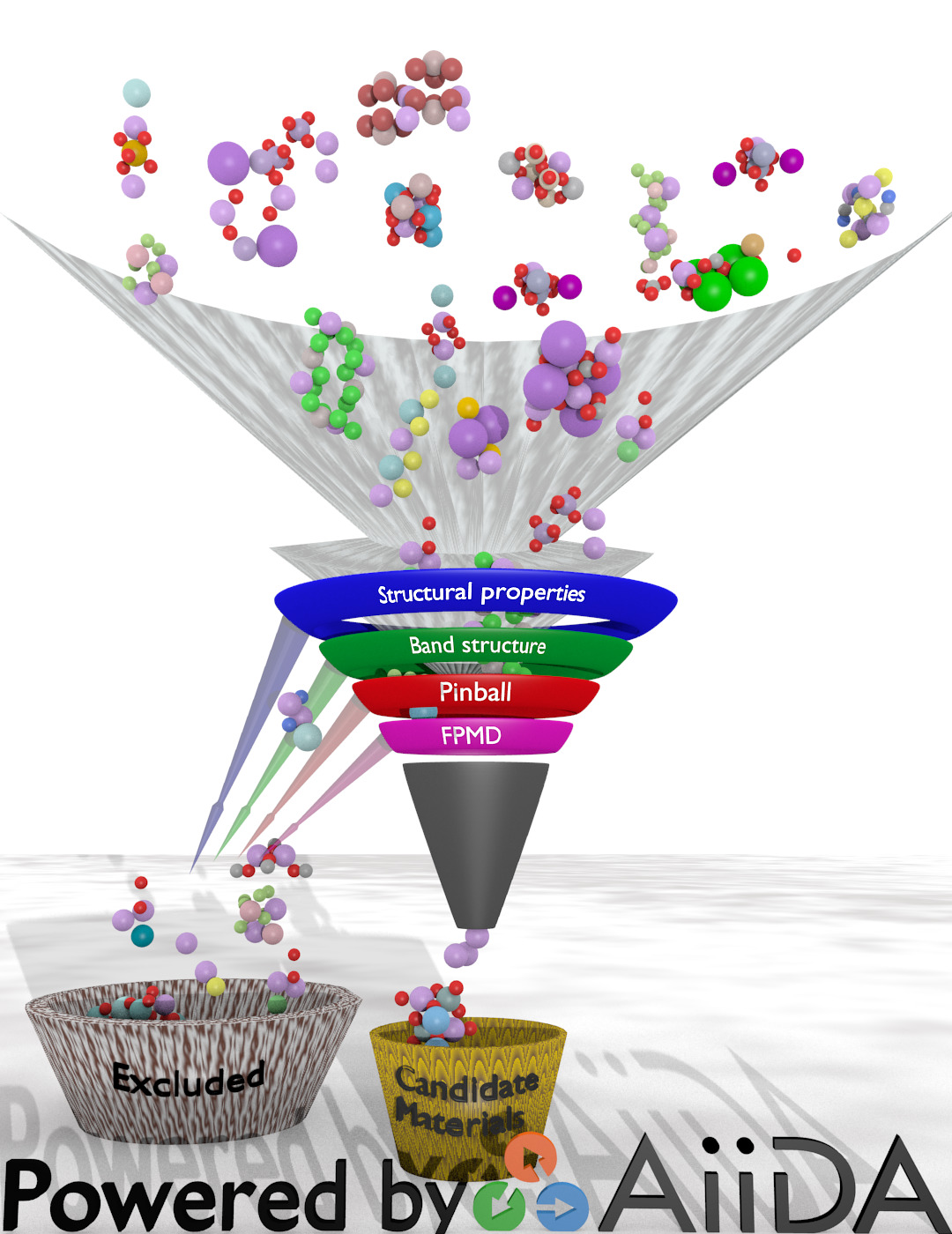}
    \caption{
    Schematic representation of the screening funnel.
    Structures downloaded from experimental repositories go sequentially through several computational filters.
    Each stage of the screening discards structures that are unsuitable as solid-state electrolytes based on ever more complex calculated properties.
    The final outcome is of a few tens of viable structures, that could be potential candidates for novel solid-state Li-ion conductors.}
    \label{fig:HT-schema}
\end{wrapfigure}

To a large extent, chemical intuition drove in the past the discovery of new solid-state ionic conductors.
As a first example, Thangadurai and Weppner~\cite{thangadurai_novel_2003} found the garnet structure to be a fast-ion conductor.
The general formula of the Li-containing garnets is \ce{Li5La3M2O12} (M=Ta, Nb)~\cite{knauth_inorganic_2009},
but substitution with aliovalent ions can increase or decrease the Li-ion concentration, resulting in a general structural formula~\cite{kozinsky_effects_2016} of Li$_x$B$_3$C$_2$O$_{12}$ (B=La, Ca, Ba, Sr, Y, ...; C=Zr, Ta, Nb, W, ...), where $x$ can vary from 3 to~7.
A second example where chemical intuition lead to a new family of superionic conductors is the recent discovery of Li-argyrodites~\cite{deiseroth_li6ps5x:_2008}, with the general formula \ce{Li7PS5X} (X=Cl, Br, I) or \ce{Li7PS6}
(sulphur can be replaced with oxygen, but this reduces the ionic conductivity~\cite{kong_li6po5br_2010}).
Third, Li-containing NASICONs (sodium superionic conductors) are phosphates with the structural formula
Li$_{1+6x}$X$_{4+2-x}$Y$_{3+x}$(PO$_4$)$_3$ (X=Ti, Ge, Hf, Zr, ...; Y=Al, Ga, Sc, Y, La, ...)~\cite{anantharamulu_wide-ranging_2011,bachman_inorganic_2016},
constituting a family of versatile compounds forming three-dimensional intercalated channels.
They originate from work that proposed a structure with suitable channels for Na$^+$-ion diffusion~\cite{hong_crystal_1976}, namely $\mathrm{Na_{1-x}Zr_2P_{3-x}Si_xO_{12}}$.
Fast Li-ionic diffusion in this family was investigated shortly after for \ce{LiZr2(PO4)3}~\cite{petit_fast_1986}, \ce{LiIn2(PO4)3}~\cite{pronin_ionic_1990}, and doped lithium-titanium phosphates~\cite{aono_ionic_1989,aono_ionic_1990}.

Chemical substitutions in known ionic conductors have also led to the discovery of new fast-ionic conductors.
The family of Li-superionic conductors (LISICON) is a widely studied group of compounds originating from the structural formula Li$_{3+x}$(P$_{1-x}$Si$_x$)O$_4$,
where Li$_3$PO$_4$ is mixed with Li$_4$SiO$_4$, changing the crystal structure from monoclinic ($x < 0.4$) to
orthorhombic ($x > 0.5$)~\cite{hu_ionic_1977,deng_structural_2015}.
Aliovalent substitutions led to the identification of fast-ionic conductors of the general chemical formula $\mathrm{Li_{4+x-z}X_x Y_{1-x-z}Z_zO_4}$, (X=B,Al,Zr,..., Y=Si,Ge,Ti,..., Z=P,As,V,...)~\cite{shannon_new_1977,rodger_li+_1985,deng_enhancing_2017}.
Substituting oxygen with sulphur 
yields the sub-family of thio-LISICONs~\cite{kanno_lithium_2001},
widely regarded as better conductors than oxygen-based LISICONs due to the higher polarizability of the S$^{2-}$ anions compared to O$^{2-}$ anions~\cite{bachman_inorganic_2016, kanno_lithium_2001}.
Additional substitutions of the cations with Ge or Sn  allow for further compositional variety within this family.
As part of the $\mathrm{Li_{4-x}Ge_{1-x}P_xS_4}$ system, tetragonal \ce{Li10GeP2S12} (LGPS), discovered by Kamaya \textit{et al.}~\cite{kamaya_lithium_2011} in 2011, is widely considered one of the current best ionic conductors.
In summary, major breakthroughs in the discovery of either new families of Li-ionic conductors or via substitutions within known families have been mostly led by chemical intuition, with simulations limited to providing new insight on diffusion mechanism in known materials.

However, the search for new solid-state electrolytes with computational methods can be highly effective.
Synthesis of ionic compounds and measurement of the ionic conductivity are labor-intensive tasks. 
In addition, experimental results can be difficult to interpret, as evidenced by the  2011 discovery of superionic tetragonal LGPS~\cite{kamaya_lithium_2011}, even though the same composition was investigated already in 2001~\cite{kanno_lithium_2001} albeit without reporting the tetragonal phase.
On the other hand, the calculation of material properties can be highly automated and parallelized~\cite{ceder_recharging_2011,hautier_novel_2011,pizzi_aiida:_2016, curtarolo_high-throughput_2013}.
A computational screening for high Li-ionic conductivity, as schematically shown in \myfigref{fig:HT-schema}, can probe new structural families for promising Li-ionic conductors, with key properties of interest being electrochemical stability and ionic diffusivity.

Electrochemical stability 
can be calculated from first principles using Kohn-Sham density functional theory~\cite{hohenberg_inhomogeneous_1964,kohn_self-consistent_1965} and grand-potential phase diagrams~\cite{ong_phase_2013,richards_interface_2016}.
As an example of computational studies addressing the question of stability and leading to new candidate structures, recent work~\cite{aykol_computational_2019} screened the garnet family of structures with $x=3$ for candidates on or close to the convex hull, resulting in 30 new structures in this family.

The diffusion of Li ions and Li-ionic conductivity can also be addressed with atomistic simulations.
Computational techniques such as molecular dynamics~\cite{alder_phase_1957,alder_studies_1959,rahman_correlations_1964} allow to predict diffusion coefficients and offer insights on the evolution of an atomic system over time in a well-defined thermodynamic ensemble.
The dynamics can be driven by
classical force fields, with parameters chosen to reproduce experiments or first-principles calculations, or using first-principles (on-the-fly) approaches based on accurate electronic-structure methods.
Historically, the garnet family, due to its large unit cells, was studied extensively using classical force field~\cite{adams_ion_2012, xu_mechanisms_2012, kozinsky_effects_2016, klenk_finite-size_2016, burbano_sparse_2016}.
Force fields were also successfully applied to the LISICON family~\cite{deng_structural_2015,deng_enhancing_2017}; 
as an example,   Adams and Rao~\cite{adams_structural_2012}  could show via classical simulations of \ce{Li10GeP2S12} that Li ions (partially) occupy an additional site that was not observed in the first synthesis of the material by Kamaya \textit{et al.}~\cite{kamaya_lithium_2011}, with experimental evidence for this site later provided by Kuhn \textit{et al.}~\cite{kuhn_single-crystal_2013}
Classical force fields have also been used when studying grain-boundary diffusion, again due to inherently large system sizes, as is the case example in Li-rich antiperovkites~\cite{dawson_atomic-scale_2018}.
On the other hand, such classical interatomic potentials have always to be carefully fitted for the application in mind.
As a consequence, they are often not general enough, and their reliability is questionable for a large-scale screening effort which implies considerable compositional variety.
Density-functional theory~\cite{kohn_self-consistent_1965} can provide an accurate and general Hamiltonian for the evolution of atoms, 
with Car-Parrinello molecular dynamics~\cite{car_unified_1985} (CPMD) allowing for a particularly efficient implementation in insulators by propagating electrons and ions together, in contrast to the conventional and more broadly applicable Born-Oppenheimer molecular dynamics (BOMD), that relies on full self-consistency of the electronic degrees of freedom at every step.
Superionic conductors were first modelled using CPMD by Cavazzoni \textit{et al.} in 1999~\cite{cavazzoni_superionic_1999}, namely superionic $\mathrm{H_2O}$ and $\mathrm{NH_3}$ at high pressure and temperature,
while in 2006 Wood and Marzari~\cite{wood_dynamical_2006} performed CPMD simulations to study the dynamics of the superionic conductor \ce{AgI} at ambient conditions.
Li-ion diffusion was modeled with BOMD only in the last decade, due to its requirement of large system sizes and long time scales~\cite{ong_phase_2013, mo_first_2012,xu_one-dimensional_2012,jalem_concerted_2013, mo_insights_2014, meier_solid-state_2014, wang_design_2015, chu_insights_2016, zhu_li3yps42_2017, marcolongo_ionic_2017}.

Many studies tried to overcome the time limitations of first-principles molecular dynamics (FPMD, to cover both BOMD and CPMD) by calculating migration barriers from static calculations~\cite{van_der_ven_first-principles_2001, du_li_2007, holzwarth_computer_2011, lepley_structures_2013,du_structures_2014, lang_lithium_2015}.
However, the collective nature~\cite{kozinsky_effects_2016, he_origin_2017, morgan_benjamin_j._lattice-geometry_2017, marcolongo_ionic_2017} of Li-ion diffusion in fast conductors makes the estimate of the  dominant transitions paths complex.
A large-scale screening via estimate of migration barriers would be cumbersome, since the definition of migration pathways remains at this stage human-intensive.
Automated calculations of pathways and barriers tackle single-particle migration~\cite{adams_bond_2002,mace_automated_2019, kahle_unsupervised_2019} in the dilute limit, thus neglecting collective effects.

Several descriptors for ionic diffusion have been suggested to circumvent the need for a direct calculation, to avoid unreliable force fields, expensive FPMD, or the identification of complex migration pathways.
A first example is the correlation between the diffusion coefficients or Arrhenius barriers of diffusion and the frequencies of specific optical phonons, first presented by Wakamura and Aniya~\cite{aniya_phonons_1996, wakamura_roles_1997, wakamura_effects_1998} for halides.
Evidence for such correlation has also been found in LISICONs, and in the family of olivines~\cite{muy_tuning_2018},
and this descriptor constituted the backbone for a very recent screening for solid-state Li-ion conductors~\cite{muy_high-throughput_2019}.
As a second example, the importance of accessible volume for diffusion~\cite{kummer_-alumina_1972, adams_bond_2002} led to the development of the bond-valence method by Adams and Swenson~\cite{adams_determining_2000}, and several efforts~\cite{avdeev_screening_2012,xiao_candidate_2015} employed this inexpensive method for screening purposes.
An approach that tackles the collective nature of diffusion is proposed via an entropy descriptor by Kweon \textit{et al.}~\cite{kweon_structural_2017} to account for site frustration.
Together with local bond frustration~\cite{adelstein_role_2016,kweon_structural_2017} and dynamical frustration given by the interaction of Li ions with a dynamically changing landscape, a picture emerges that explains fast-ionic diffusion in the class of \textit{closo-}borates.
In addition, machine-learning tools have been applied to predict ionic diffusion from a manifold of possible descriptors~\cite{d.sendek_holistic_2017}, but with limited data available, training machine-learning models to the required accuracy is ambitious.
We conclude this cursory overview of computational methods with the comment that screening for fast-ion conductors with existing methods remains challenging, either due to the limited accuracy of descriptors or force-fields, or the complexity and computational cost of first-principles approaches.

Calculating the diffusion of Li ions in a screening scenario accurately, with molecular dynamics, calls for a different approach, that combines the computational efficiency of force fields with the generality and accuracy of density-functional theory.
We recently proposed the pinball model~\cite{kahle_modeling_2018} as an accurate framework for the dynamics of Li ions in the solid state.
Details and derivation are in the original paper, but we iterate here the key assumptions behind the model.
First, since Li ions negligibly perturb the valence electronic charge density of an ionic system, the charge density is assumed to no longer depend on the instantaneous positions of Li ions.
Second, since freezing the host lattice (i.e. everything other than lithium) has a minor effect on the resulting dynamics (especially for stiff systems), we consider the Li ions moving in a frozen host lattice with a frozen charge density.
We use this pinball model as the backbone for a high-throughput screening effort to find fast ionic conductors, to be followed by extensive first-principles simulations for the most promising candidates.
We present details of the methods in \mysecref{sec-methods}, also laying out the automatization efforts undertaken.
This is followed by presentation and discussion of the results in \mysecref{sec-results}.
We summarize and give our conclusions and outlook in \mysecref{sec-conclusions}.

\section{Methods}
\label{sec-methods}

\subsection{Automation and provenance}
\begin{figure*}
\centering
\includegraphics[width=1.0\hsize]{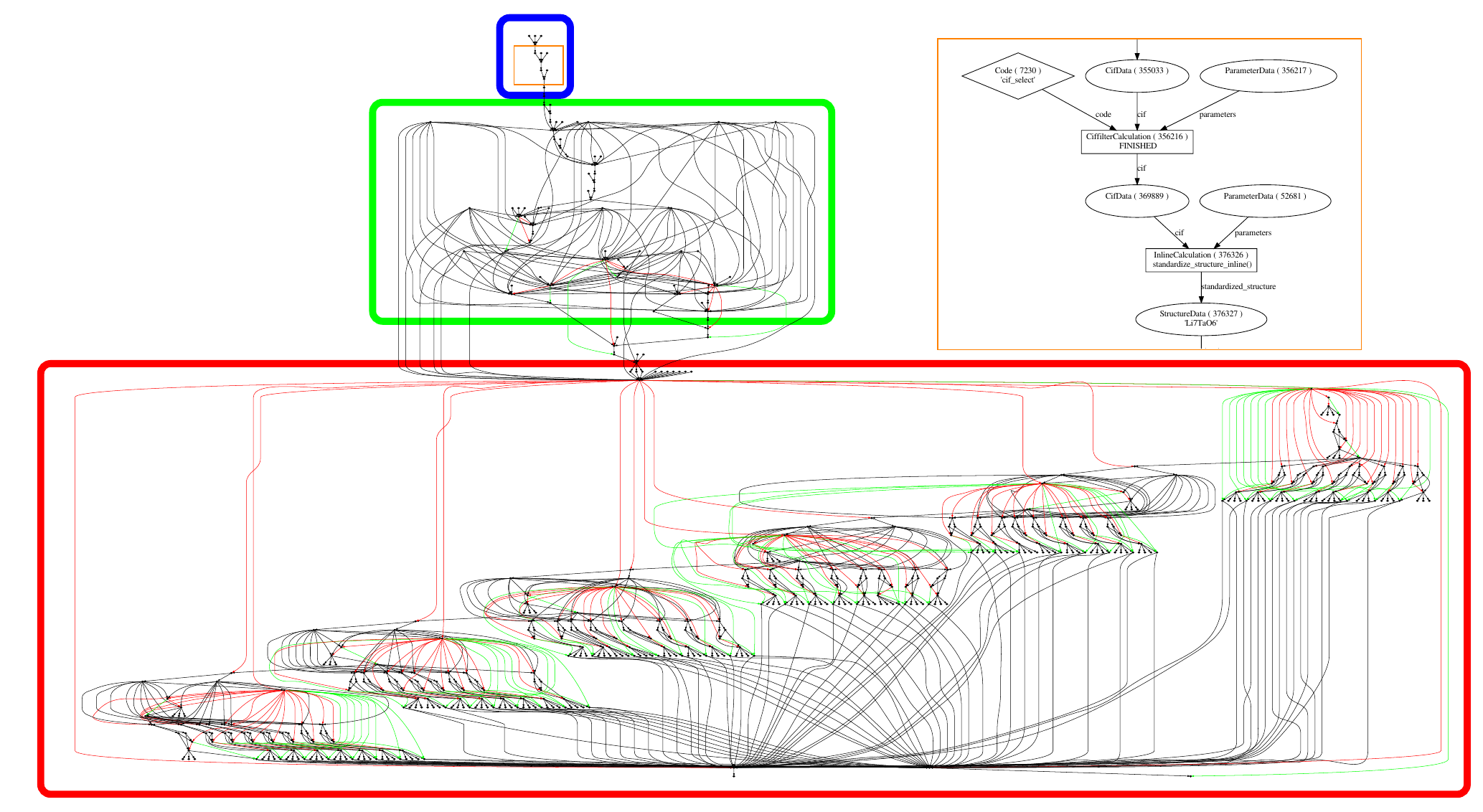}
\caption{
The storage of every calculation and data-instance in a directly acyclic graph in AiiDA is shown above.
Every node in the graph is a calculation or data instance, with black lines denoting data provenance (a calculation creating data, or a data instance being an input to a calculation), green lines denoting logical provenance (a workflow returning a data instance) and red lines referring to operational provenance (a workflow calling a workflow or calculation).
The subgraph corresponding to the structure ingestion is contained inside the blue rectangle; the subsequent calculations of band structures, variable-cell relaxations, and fitting of the pinball model are inside the green rectangle, while the calculation of the diffusion coefficient with the pinball model at one temperature is shown in the red rectangle.
The orange inset enlarges one region of the graph, where the data provenance is given in greater detail for the standardization of the \ce{Li7TaO6}  structure during structure ingestion; in this inset, codes are given by diamonds, data objects by circles, and calculations by squares.
}

\label{fig:pw-calc-graph}
\end{figure*}

Any high-throughput effort requires a highly automated framework for launching, monitoring, parsing, and storing a large number of calculations on many structures.
Recording explicitly, ideally in an easily queryable format, the provenance of the resulting data allows for fully reproducible results~\cite{ceder_opportunities_2010,hautier_computer_2012,curtarolo_high-throughput_2013,hautier_finding_2019,alberi_2019_2018}.
To achieve automation and explicit storage of the provenance, we leverage the Automated Interactive Infrastructure and Database for Computational Science (AiiDA) materials informatics platform, developed by Pizzi \textit{et al.}~\cite{pizzi_aiida:_2016}
The novelty of AiiDA in the field of materials informatics is that every calculation is stored as a node in a graph, with input data forming incoming nodes, and output data stored as outcoming nodes, that can again be input to a different calculation.
To illustrate the principle, we show such a graph (from the database created in this work) for a single structure in \myfigref{fig:pw-calc-graph}.
The resulting directed acyclic graph stores the full provenance of every result. 
In addition, AiiDA allows for a high degree of automation and parallelization via its daemon.
Every calculation presented in this work is run with AiiDA.

\subsection{Structure ingestion}

CIF files of all Li-containing structures are retrieved from two structural repositories, namely the Inorganic Crystal Structure Database (ICSD)~\cite{belsky_new_2002} and the Crystallography Open Database (COD)~\cite{grazulis_crystallography_2012}, using tools that are provided by AiiDA.
Here, we disregard structures that have partial occupancies or attached hydrogen.
This is mainly due to the complexity of creating different derivative configurations, that usually require sampling strategies~\cite{van_de_walle_alloy_2002,hart_algorithm_2008}.
The additional refinement of CIF-files is described in the work by Mounet \textit{et al.}~\cite{mounet_two-dimensional_2018}
We employ the same protocol, using COD-tools~\cite{merkys_cod::cif::parser:_2016} to standardize the CIF-files, and
the structure-matcher of pymatgen~\cite{ong_python_2013} to compare crystal structures using the CMPZ-algorithm~\cite{hundt_cmpz_2006} in order to remove duplicates and work with unique structures.
Parameters used and additional details are given in \appsecref{app-meth-duplicates}.

\subsection{Structural properties}
Additional filters are applied to exclude certain elements:
hydrogen-containing compounds (because the effect of light hydrogen on Li motion in the pinball model has not been studied), and elements
that are very rare, or dangerous (details given in \appsecref{app-meth-composition-filters}).
An additional filter is applied on atomic distances to exclude common organic compounds and structures where atomic distances are so small that we have to assume a corrupted or incorrect representation (additional details given in \appsecref{app-meth-bond-filters}).

\subsection{Electronic structure}
\label{sec-electronic}
In order to estimate whether a structure is electronically insulating, we perform a single SCF calculation at the experimental geometry using density-functional theory.
We perform all DFT simulations in this work with the pw.x code, part of the Quantum ESPRESSO distribution~\cite{giannozzi_quantum_2009}, and use PBE~\cite{perdew_generalized_1996} as the exchange-correlation functional.
We always take pseudopotentials and cutoffs from the Standard Solid-State Pseudopotential (SSSP) Efficiency 1.0 library~\cite{prandini_precision_2018}, that verifies pseudopotentials from different methods and libraries~\cite{willand_norm-conserving_2013,dal_corso_pseudopotentials_2014,garrity_pseudopotentials_2014,topsakal_accurate_2014,schlipf_optimization_2015,van_setten_pseudodojo:_2018}.
For this initial estimate of the electronic structure we use Marzari-Vanderbilt cold smearing~\cite{marzari_thermal_1999} (additional details given in \appsecref{app-meth-electronic}). A system is judged as insulating if the lowest valence state shows negligible electronic occupation (see \appsecref{app-meth-electronic}), which is a function of the band gap. This is generally underestimated by PBE, but the criterion we chose is not too strict. 
For all insulating structures, we proceed with a variable-cell relaxation to the ground-state geometry, as explained in \appsecref{app-meth-vc}.

\subsection{Diffusion in the pinball model}
\label{subsec-diff-pinball}

Supercells for molecular dynamics simulations are created from every relaxed structure as specified in \appsecref{app-supercells}, with a minimum distance criterion $d_{inner} = 8$~\r{A} between opposite faces.
The Hamiltonian of the pinball model~\cite{kahle_modeling_2018} is:
\begin{align}
    \mathcal{H}_P =&  \frac{1}{2} \sum^P_p M_p \dot{\bm R}_p^2 + \alpha_1 E_N^{P-P} + \alpha_2 E_N^{H-P} 
	+ \beta_1 \sum_p^P \int n_{R_{H_0}}(\bm r) V_p^{LOC}(\bm r)  d\bm r,
    \label{eq.pinball-screened}
\end{align}
where $\bm R_p$ are the positions of pinballs (i.e. the Li ions) and $\dot{\bm R}_p$ their velocities;  $E_N^{P-P}$ is the electrostatic interaction between the pseudopotentials cores of the pinballs,  $E_N^{H-P}$ the interaction between the pinballs and the host lattice, $V_p^{LOC}$ the local pseudopotential of a Li core and its 1s electrons, and $n_{R_{H_0}}(\bm r)$ is the frozen charge density of the system that parametrically depends only on the ground state positions of the host lattice $\bm R_{H_0}$.
With respect to the original formulation~\cite{kahle_modeling_2018}, we neglect the non-local interactions of the pinball pseudopotential core with the frozen wavefunctions of the system.
As discussed in the same reference, this term adds some accuracy, but comes at a higher computational cost.
The local pinball of \myeqref{eq.pinball-screened} used in this work has only quadratic scaling with system size, compared to cubic scaling when including non-local interactions.
For all supercells with up to 500 atoms we fit the parameters $\alpha_{1}$, $\alpha_{2}$ and $\beta_1$ for the pinball Hamiltonian, with details for the fitting given in \appsecref{app-fitting-pinball}.
Larger supercells are discarded from the screening, due to the computational costs.

All structures that are successfully fitted are passed to the diffusion workflow,
that converges the diffusion coefficient at a given temperature to a desired threshold.
The target temperature is 1000~K, and independent blocks sampling the canonical ensemble (constant number of particles, volume and temperature) are obtained as specified in \appsecref{app-temperature-control}.
For each block, the tracer diffusion coefficient of Li $D_{tr}^\mathrm{Li}$ is calculated from the mean-square displacement as:
\begin{align}
D_{tr}^\mathrm{Li}=& 
\lim_{t \rightarrow \infty} \frac{1}{6t}  \left\langle  \mathrm{MSD}(t) \right\rangle_{NVT}
=\lim_{t \rightarrow \infty}   \frac{1}{6t}\frac{1}{N_\mathrm{Li}} \sum_l^\mathrm{Li}  \left\langle |\bm R_l(t+\tau)-\bm R_l(\tau)|^2 \right\rangle_\tau,
\label{eq-Dtr}
\end{align}
where $\langle\cdots\rangle_{NVT}$ indicates the average over the canonical ensemble sampled ergodically by the molecular dynamics simulation, replacing thus the ensemble average with a time average $\langle\cdots\rangle_\tau$.
We fit the slope of $\mathrm{MSD}(t)$ between 8~ps and 10~ps. 
For all analysis of the trajectories, we use tools of our open-source Suite for Analysis of Molecular Simulations (SAMOS)~\cite{lekah_suite_2019}.

\subsection{First-principles molecular dynamics}
For structures showing significant diffusion in the pinball model at 1000~K, we calculate the diffusion coefficient at the same temperature with FPMD.
However, we exclude structures that are referred to as unstable in the experimental reference entered into the databases, or that are already well-known ionic conductors, in order to focus the computational time on systems that are not being studied actively in the literature,
since the main purpose of this work is screening for novel ionic conductors.
We perform Born-Oppenheimer molecular dynamics, with details given in \appsecref{app-meth-fpmd}, and estimate the diffusion coefficient also using \myeqref{eq-Dtr}.

For structures that show significant diffusion in the FPMD simulations at 1000~K, we calculate the diffusion coefficient, using the same parameters, at three lower temperatures, namely 750~K, 600~K, and 500~K, which are equidistant on the inverse temperature scale of the Arrhenius plot.
For the structures that show significant diffusion also at the lowest temperature, we estimate the barrier to diffusion from a linear fit to the Arrhenius behavior.
For several structures, we calculate the Li-ion (probability) density $n_{Li}({\bm r})$ to visualize connected diffusive components in the system:
\begin{equation}
 n_{Li}(\bm r) =\left\langle \sum_l^{Li} \delta(\bm r-\bm R_l(t)) \right\rangle_t,
 \label{eq-density}
\end{equation}
where $l$ runs over the Li ions in the system, whose positions at time $t$ are given by $\bm R_l(t)$, and the angular brackets $\langle\cdots\rangle_t$ indicate a time/trajectory average.
In practice, we replace  the delta function by a Gaussian with a standard deviation of 0.3~\r{A}, and perform the summation on a grid (of 10~points per \r{A} in every direction), as implemented in SAMOS~\cite{lekah_suite_2019}.
To show Li-ion densities, we plot isosurfaces at values of 0.1, 0.01, and 0.001~\r{A}$^{-3}$  in cyan, blue, and purple, respectively.

\section{Results and discussion}
\label{sec-results}

\begin{wrapfigure}{R}{0.5\hsize}
\includegraphics[width=\hsize]{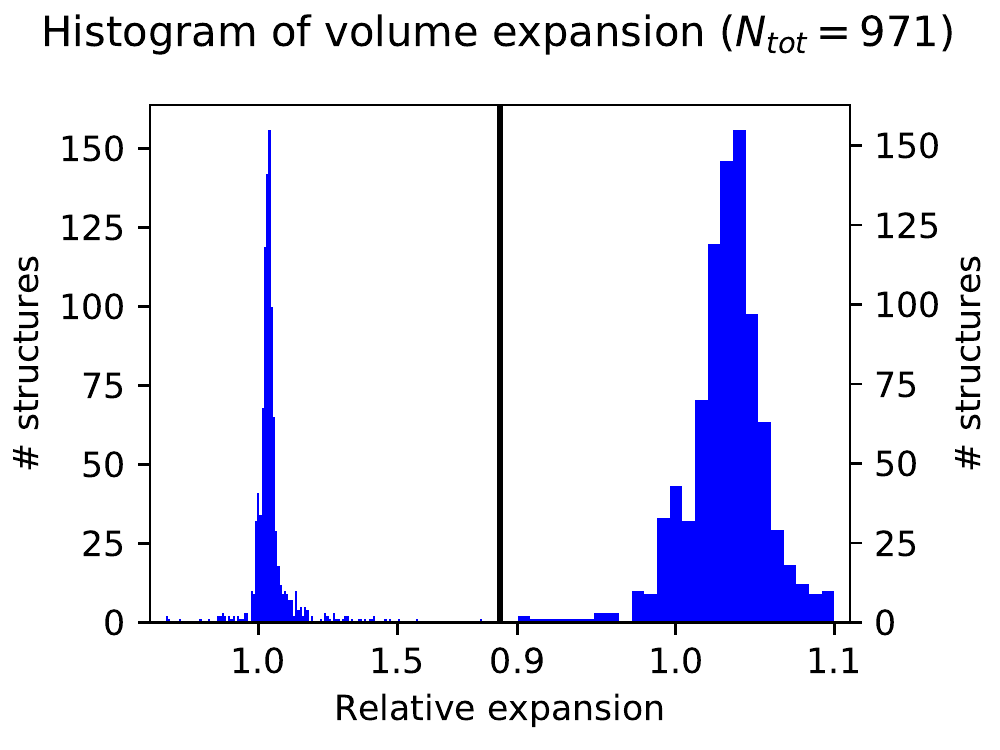}
\caption{
Histogram of the relative volume expansion between the experimental and the calculated ground-state volumes (at the PBE level) for 971 structures studied.
In the left panel, we show the entire histogram, on the right we zoom on the region between 0.9 and 1.1.
}
\label{fig-volume-expansion}
\end{wrapfigure}

Downloading all Li-containing compounds in the ICSD~\cite{belsky_new_2002} and COD~\cite{grazulis_crystallography_2012} structural repositories results in 8627 and 7228 entries, respectively.
The first filter ensures that structures have no attached hydrogens and no partial occupancies, and leads to 3956  and 3777  structures from the ICSD and COD, respectively.
We extract a total of 7472 valid structures (261 CIF-files could not be interpreted by the pymatgen~\cite{ong_python_2013} CIF-reader) from the CIF-files, of which we find 4963 to be unique using the pymatgen structure matcher.
1362 of these unique structures pass also the filters on elements and bond distances (see Secs.~\ref{app-meth-composition-filters} and \ref{app-meth-bond-filters}), and of these 1016 are insulators at the PBE-DFT level, according to the criterion explained in \mysecref{app-meth-electronic}.
971 structures could be successfully relaxed, the remainder failing due
problems with the iterative self-consistency.
A histogram of the volumes after relaxation divided by the volume before relaxation (i.e. the experimental volume from the database), as shown in \myfigref{fig-volume-expansion}, reveals that a structure is more likely to expand than contract, as expected for the PBE functional.
While the volume changes are rather small (peaked at 4\%, i.e. 1.3\% per direction for isotropic expansion) for almost all cases, there are outliers that expand or contract substantially,
likely due to  van der Waals interactions (for example, layered materials).

We were able to obtain the coefficients $\alpha_1$, $\alpha_2$,  and $\beta_1$ of \myeqref{eq.pinball-screened} in 916 cases, of which 903 are judged to be sufficiently good based on the $r^2$ correlation between DFT and pinball forces.
Failure to fit in the remaining cases is mainly due to failures of the iterative self-consistent convergence for training configurations.
The diffusion workflows in the pinball model completed successfully in 796 cases, summing up to a simulation time of 7.6~$\mu$s.
Reasons for this 12\% failure rate are drifts in the constant of motion and the inability to converge the diffusion coefficient.
We take the 200 systems with the highest diffusion in the pinball model at 1000~K, and  exclude from this set systems that either have been reported as good ionic conductors in the literature and have been studied independently by several groups, or that are referred to as unstable at room temperature in the experimental reference.

Before discussing the results of FPMD, we give an overview of the structures that we exclude because they have been already studied substantially in the literature.
The classification of all know ionic conductors found at this stage of the screening allows for an assessment whether the screening can be considered holistic, meaning that the entire compositional variety is captured.
Known LISICON structures that we find are  \ce{Li7P3S11}~\cite{chu_insights_2016}, 
\ce{Li4GeS4}~\cite{murayama_structure_2002}, and \ce{Li4SnS4}~\cite{choi_coatable_2017} from the LISICON family, 
and \ce{Li5La3Ta2O12} and \ce{Li5La3Nb2O12}~\cite{thangadurai_novel_2003, allen_effect_2012,logeat_order_2012,xu_mechanisms_2012} from the garnet family.
We miss the prototypical garnet structure \ce{Li7La3Zr2O12} because of failures (drifts of the constant of motion) during the pinball dynamics.
We find also many NASICONs such as 
\ce{Li3Sc2P3O12}~\cite{novoselov_phase_2008},  \ce{Li3In2P3O12}~\cite{pronin_ionic_1990},  \ce{LiZr2P3O12}~\cite{petit_fast_1986}, 
\ce{LiTi2P3O12}~\cite{aatiq_structure_2002}, and \ce{Li4ZnP2O8}~\cite{saha_polymorphism_2018}.
We find in the screening both oxide and sulphide argyrodytes, namely \ce{Li6PS5I}~\cite{kong_lithium_2010}, \ce{Li6PClO5} and  \ce{Li6PBrO5}~\cite{kong_li6po5br_2010}, and \ce{Li5PS4Cl2}~\cite{zhu_li3yps42_2017}.


In the remainder, we focus on the remaining 132 materials that were studied with accurate FPMD, summing up to a simulation time of 45~ns.
We divide the analysis of candidates into four categories, based on the observed diffusion in FPMD:
(A) Structures that show diffusion at high (1000~K) and at low (500~K) temperature are classified as fast-ion conductors; (B) Structures that show diffusion at high temperature, but either show slow diffusion at lower temperatures, or could not have their diffusion resolved at lower temperature, are classified as potential ionic conductors.
(C) Structures that show negligible diffusion at high temperature are classified as not being ionically conducting;
(D) Structures where, due to computational difficulties, we can make no precise statements, but that could be good candidates based on the results of the simulations of the pinball model alone.

\subsection{Fast-ionic conductors}

\begin{wrapfigure}{R}{0.5\hsize}
\includegraphics[width=\hsize]{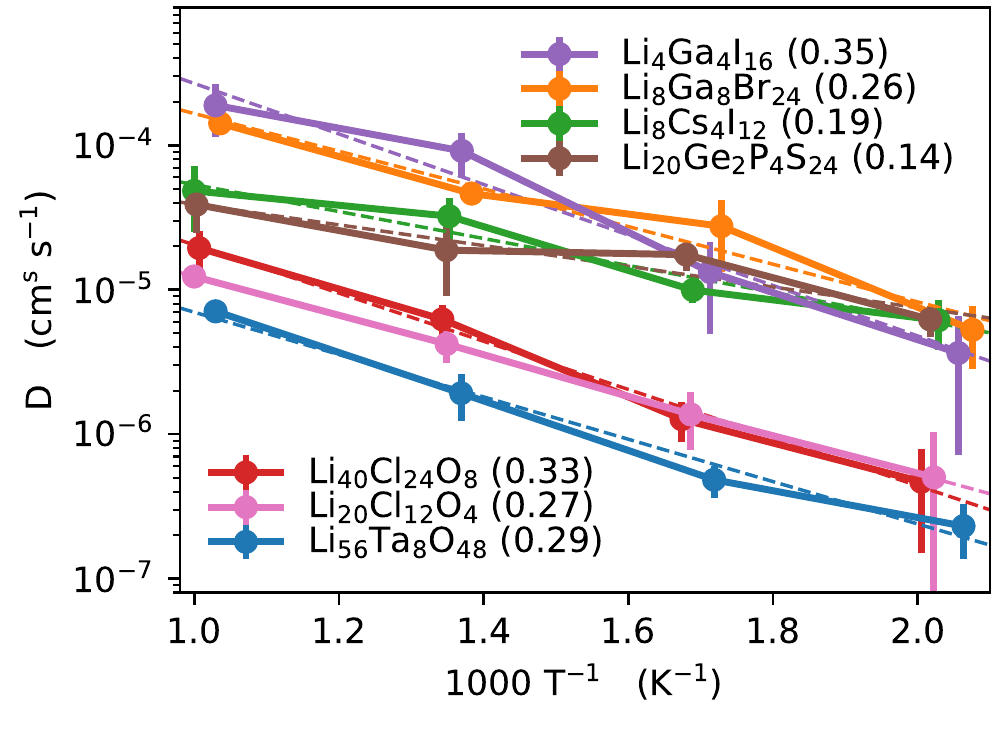}
\caption{Diffusion from FPMD for \ce{Li4Ga4I16}, \ce{Li7Ga8Br24}. \ce{Li8Cs4I12}, \ce{Li20Ge2P4S24}, \ce{Li40Cl24O8}, \ce{Li20Cl12O4}, and \ce{Li56Ta8O48} as solid lines
in orange, brown, blue, violet, green, pink, and red, respectively (we refer to the composition of the actual supercell studied).
The line of best fit is shown as a dashed line of the same color.
The activation barriers, extracted from the slope of the fit, are in the legend (in brackets) in~eV.}
\label{fig-activation-group-A}
\end{wrapfigure}

This first group includes ionic conductors that could be of significant interest for application as solid-state electrolytes.
Due to their fast conduction, we are able to resolve the diffusion also at lower temperatures and extract the activation barriers, which are shown in \myfigref{fig-activation-group-A}.
To our knowledge, they have been not been studied  extensively  or at all by experiments, apart from LGPS.
The provenance (ICSD/COD entries), volume change during cell relaxation, and simulations times of the candidates in this group are given in \suppltabref{tab-provenance-group-A}.

\begin{figure}[t]
    \centering
    \includegraphics[width=0.48\hsize]{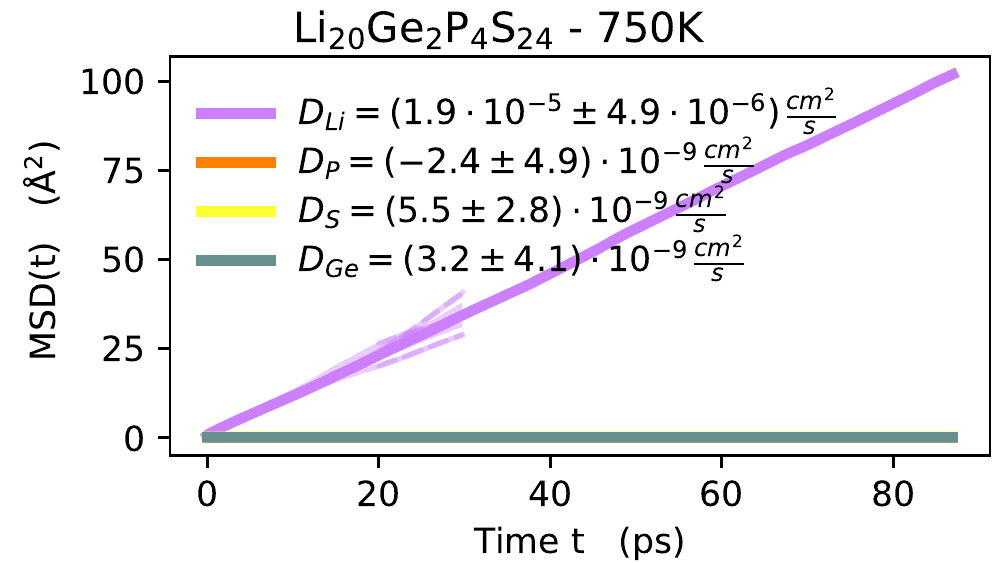}
    \includegraphics[width=0.48\hsize]{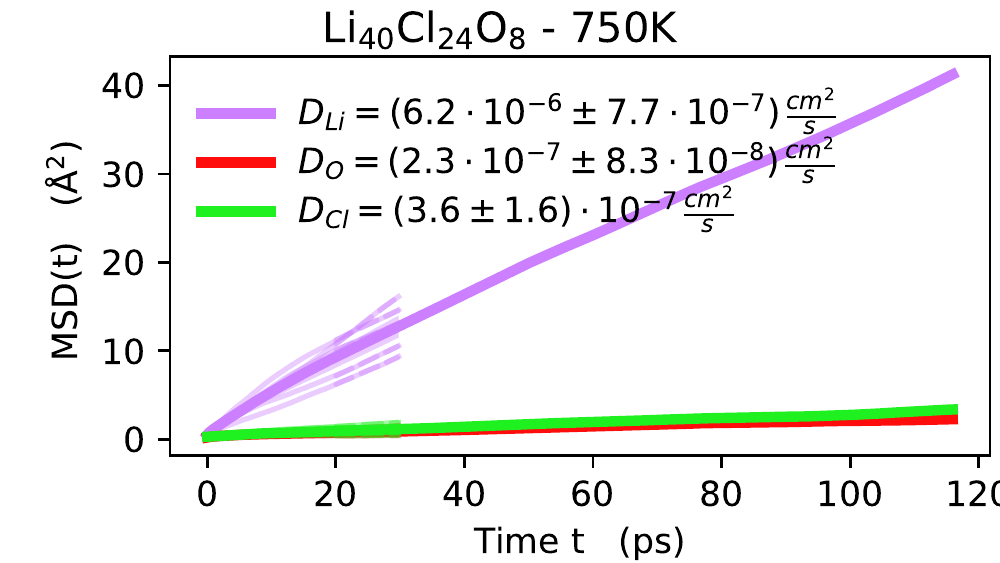}\\
    \includegraphics[width=0.48\hsize]{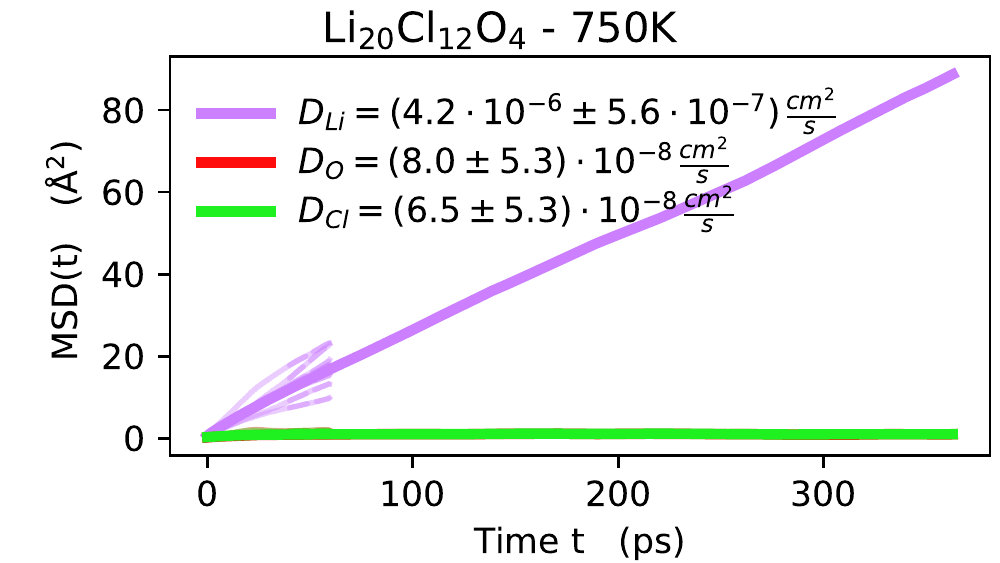}
	\includegraphics[width=0.48\hsize]{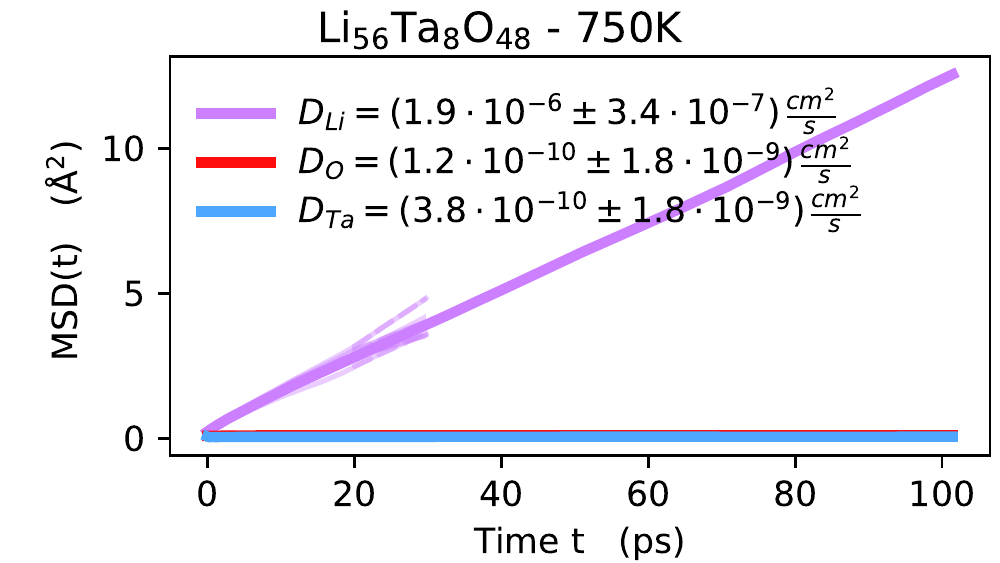}\\
    \includegraphics[width=0.48\hsize]{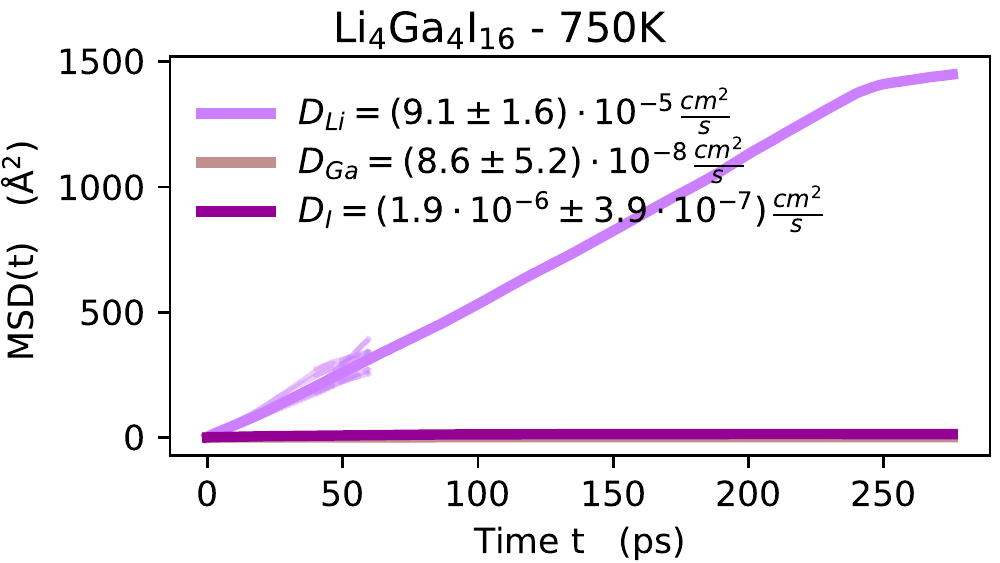}
    \includegraphics[width=0.48\hsize]{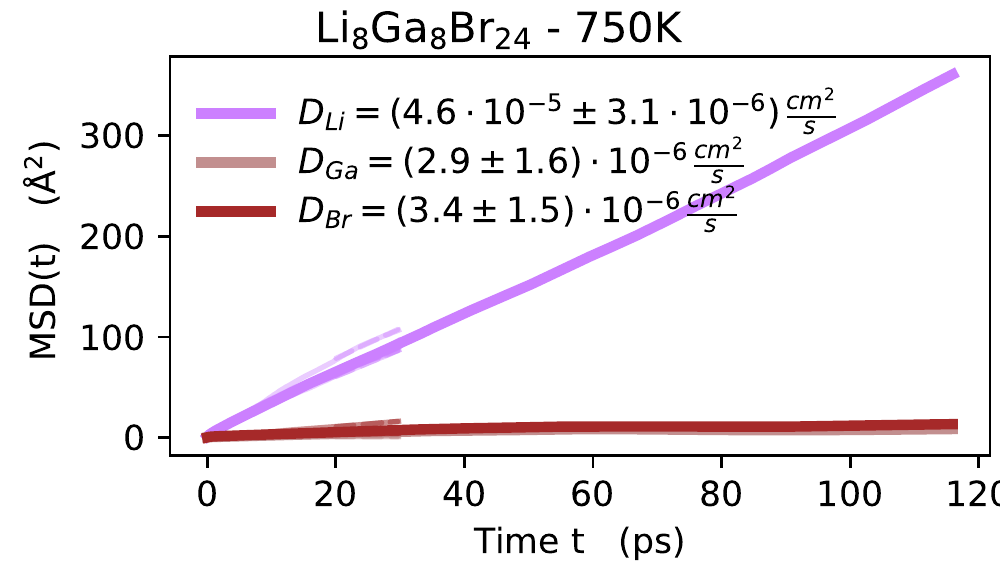}
    \caption{
We show the MSD of all species  at 750~K from FPMD in \ce{Li20Ge2P4S24}, \ce{Li40Cl24O8}, \ce{Li20Cl12O4},
 \ce{Li56Ta8O48}, \ce{Li4Ga4I16}, and \ce{Li8Ga8Br24}  (left to right and top to bottom).}
    \label{fig-msds-A1}
\end{figure}

\begin{figure}[t]
    \includegraphics[width=0.35\hsize]{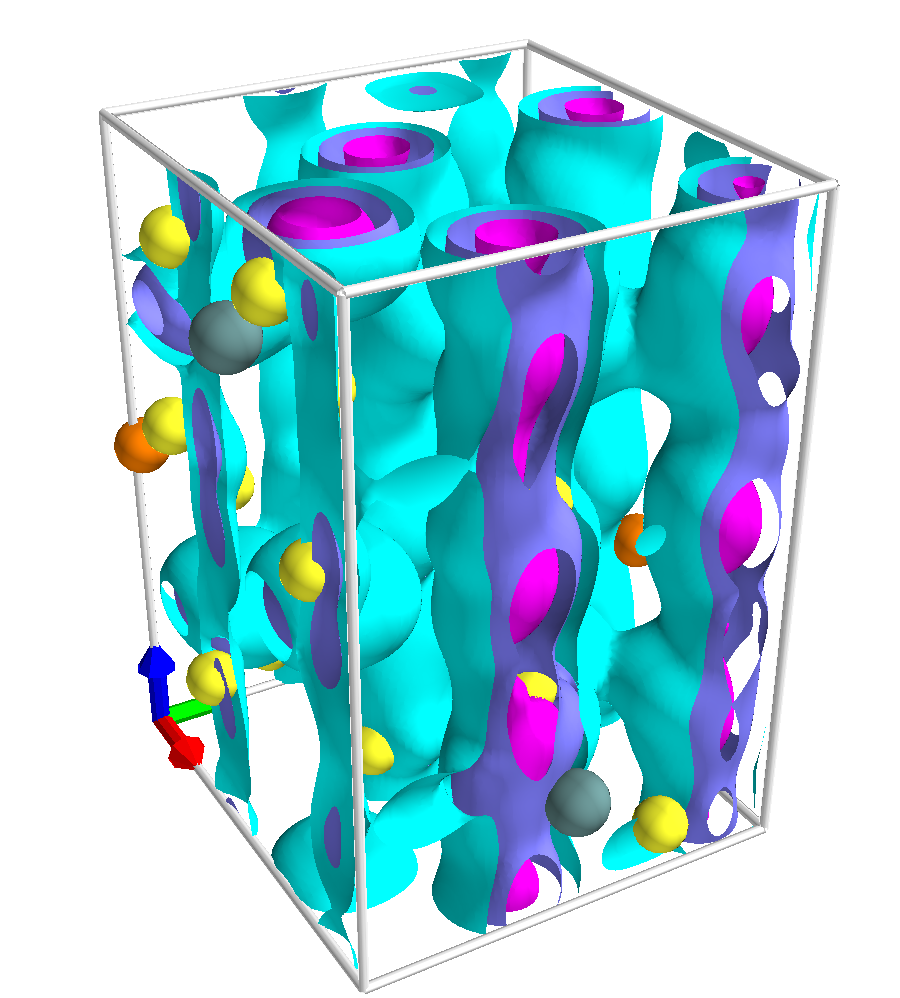}
     \includegraphics[width=0.63\hsize]{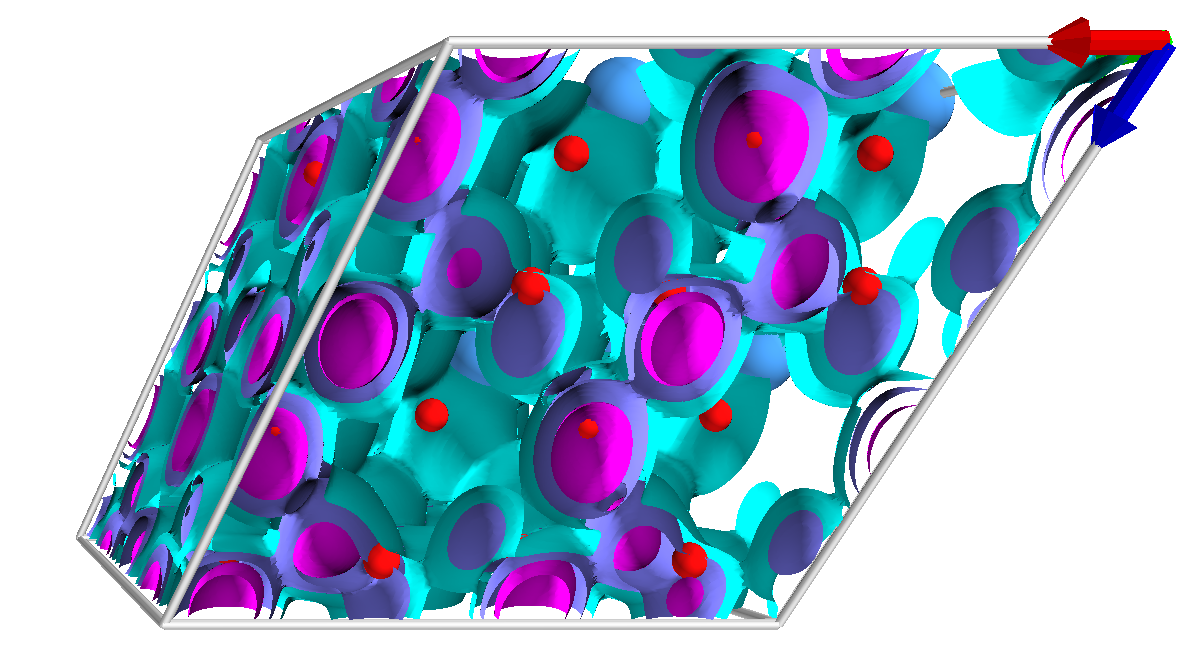}
    \caption{
    (Left) Li-ionic density of \ce{Li20Ge2P4S24} at 500~K from FPMD. The unidimensional channels along the $c$-axis (along blue arrow) are clearly visible;
    see text for detail on how we calculate the density and the isosurface levels.
    (Right) Li-ionic density of \ce{Li56Ta8O48}  at 500~K from FPMD.}
    \label{fig-dens-Li20Ge2P4S24-500}
\end{figure}

\paragraph*{\ce{Li20Ge2P4S24}:}
The well-known superionic conductor \ce{Li20Ge2P4S24}~\cite{kamaya_lithium_2011, mo_first_2012,ong_phase_2013,xu_one-dimensional_2012,marcolongo_ionic_2017} (LGPS) is included by us in the set of candidates as a reference, since it constitutes one of the best Li-ion conductors.
The MSDs (shown in \appfigref{fig-msd-Li20Ge2P4S24}) are compatible with fast-ion diffusion at every temperature studied. We show the MSDs extracted at 750~K in the top left panel of \myfigref{fig-msds-A1}.
We find an activation barrier of 0.14~eV, which is slightly lower than previous computational studies, but certainly within the error due to finite statistics in FPMD:
Marcolongo and Marzari~\cite{marcolongo_ionic_2017} estimate a value of 0.18~eV, Ong \textit{et al.}~\cite{ong_phase_2013} of 0.21~eV.
We can reproduce the strongly unidimensional conduction pathway of this material, also evident from the Li-ion density in the left panel of \myfigref{fig-dens-Li20Ge2P4S24-500}.

\paragraph*{\ce{Li5Cl3O}:}
We study this material at two different supercell sizes to control for finite size effects, namely \ce{Li40Cl24O8} and \ce{Li20Cl12O4}. 
The structure was first reported by Reckeweg \textit{et al.}~\cite{reckeweg_olaf_li5ocl3_2012},
and its usage as a SSE is speculated in the reference, but we found no evidence that this was ever tested.
\ce{Li20Cl12O4}/\ce{Li40Cl24O8} are simulated with FPMD at 500~K for 726~ps/581~ps, at 600~K for 726~ps/523~ps, at 750~K for 726~ps/232~ps, and  for 218~ps/262~ps at 1000~K.
A certain degree of host-lattice diffusion is observed, which could indicate an instability of the lattice at the increased temperatures of simulation. 
The MSD of every species at 750~K are shown in the top right and center left panels of \myfigref{fig-msds-A1}, while the other temperatures can be found in \myfigref[Figs.~]{fig-msd-Li40Cl24O8} and \supplref[]{fig-msd-Li20Cl12O4}.
The Arrhenius behavior is plotted in \myfigref{fig-activation-group-A} with the other candidates of this group.
We estimate the barriers as 0.29~eV and 0.27~eV for the larger and smaller supercell, respectively. 
The diffusion coefficients are compatible for the two supercells, indicating that our results are not subject to large finite-size effects.
The barrier is certainly low enough to classify this material as a candidate solid-state electrolyte.
However, the synthesis of the material involves elemental lithium, which means that producing a purely ionic sample without mtextit{et al.}lic side phases could be challenging for experimental validation.

\paragraph*{\ce{Li7TaO6}:}
This Li-tantalate   is
studied in its supercell \ce{Li56Ta8O48}.
Its ionic conductivity was studied experimentally by Delmas \textit{et al.}~\cite{delmas_conducteurs_1979}, Nomura and Greenblatt~\cite{nomura_ionic_1984}, and by M\"uhle \textit{et al.}~\cite{muhle_new_2004}, showing a high activation barrier of $\mathrm{0.66~eV-0.67~eV}$. The latter two references also observe a regime of low energy barrier, at low temperatures ($<50~^\circ$C) for Nomura and Greenblatt, and at high temperatures ($<400~^\circ$C) for M\"uhle \textit{et al.} 
Our simulations of 73~ps, 203~ps, 640~ps and 552~ps at 1000~K, 750~K, 600~K and 500~K, respectively, show fast-ionic diffusion within a stable host lattice and result in an activation barrier of 0.29~eV.
The MSDs for all species extracted from the simulation at 750~K are shown in the center right panel of \myfigref{fig-msds-A1} (all temperatures plotted in \appfigref{fig-msd-Li56Ta8O48}).
We plot three isosurfaces of the Li-ion densities, calculated from the lowest temperature (500~K) simulation,  in the right panel of \myfigref{fig-dens-Li20Ge2P4S24-500}, giving evidence for three-dimensional diffusion and a single connected component of diffusion.
Although the material has been studied experimentally, the high ionic diffusion from FPMD calls for additional studies for its application as a solid-state electrolyte. The substitution of Ta with aliovalent dopants could change the Li-ion concentration and improve the Li-ionic conductivity, as shown in the garnet structure~\cite{kozinsky_effects_2016}.

\paragraph*{\ce{LiGaI4} and \ce{LiGaBr3}:}
We also find the Ga-doped halides \ce{LiGaI4} and \ce{LiGaBr3} could show very promising Li-ion diffusion.
\ce{LiGaBr3} was first synthesized by H\"onle and Simon~\cite{honle_darstellung_2014}, together with \ce{LiGaBr4}, evidence that \ce{Ga} can change oxidation states in the structure. 
The same reference reports this material as a layered structure $\mathrm{Li_{2}^+[Ga_2Br_6]^{2-}}$.
\ce{LiGaI4} was synthesized by H\"onle \textit{et al.}~\cite{honle_darstellung_2014-1}
The volume expansion of 19.5\% and 14.2\% during the variable-cell relaxation hints at an instability of the structure, likely due to the lack of van-der-Waals dispersion corrections, and imply that our results from the FPMD should be interpreted with care, and that such interactions should be included in future screenings.
We plot the MSD for the supercells  \ce{Li4Ga4I16} and \ce{Li8Ga8Br24} at 750~K in the bottom panels of \myfigref{fig-msds-A1} (other temperatures are in \myfigref[Figs.~]{fig-msd-Li8Ga8Br24} and \supplref[]{fig-msd-Li4Ga4I16}).
We observe high ionic diffusion of Li ions, but also non-negligible diffusion of the sublattice, further signs of instabilities in this material.
Since indium-doped \ce{LiBr}, \ce{Li3InBr6} is a known ionic conductor~\cite{adelstein_role_2016}, our results suggest studying in greater detail also the Ga-doped Li bromide and Li iodide.

\paragraph*{\ce{Li2CsI3}:} The Cs-doped Li-iodide \ce{Li2CsI3} was first synthesized in 1983~\cite{meyer_synthesis_1983} in monoclinic phase studied by us.
In this structure, we observe a volume contraction by 13.1\%. Our simulations of 160~ps at 1000~K, 349~ps at 750~K and 726~ps at 500~K and 600~K reveal a high diffusion for Li ions, but also significant diffusion of the host lattice (see \appfigref{fig-msd-Li8Cs4I12}). It remains to be studied with different methods whether the host lattice is stable at room temperature, and the host-lattice instability is an artifact of our simulations.
Nevertheless, the high Li-ion diffusion from FPMD make this material another candidate solid-state electrolyte.

\subsection{Potential fast-ionic conductors}

\begin{figure}[t]
    \centering
    \includegraphics[width=0.48\hsize]{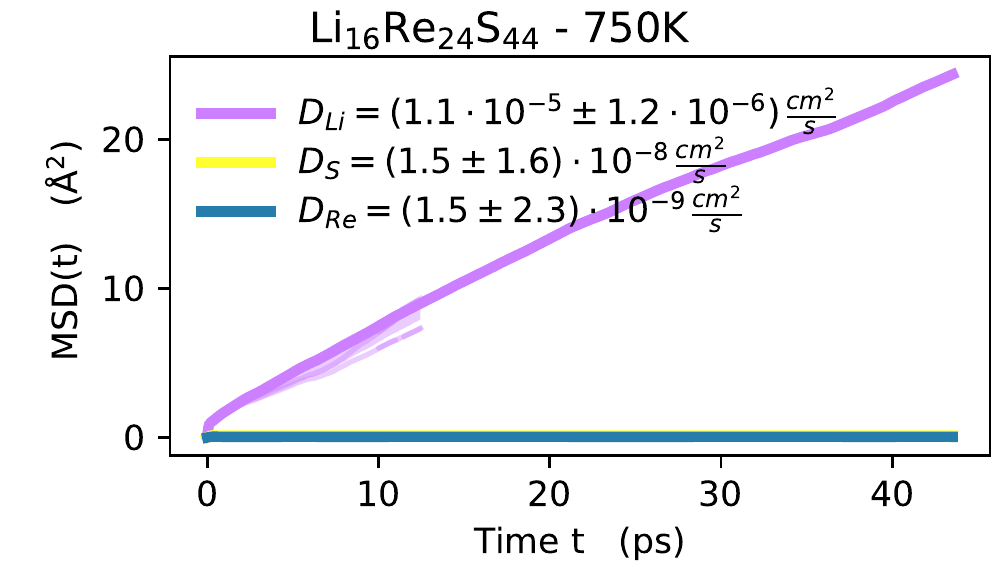}
    \includegraphics[width=0.48\hsize]{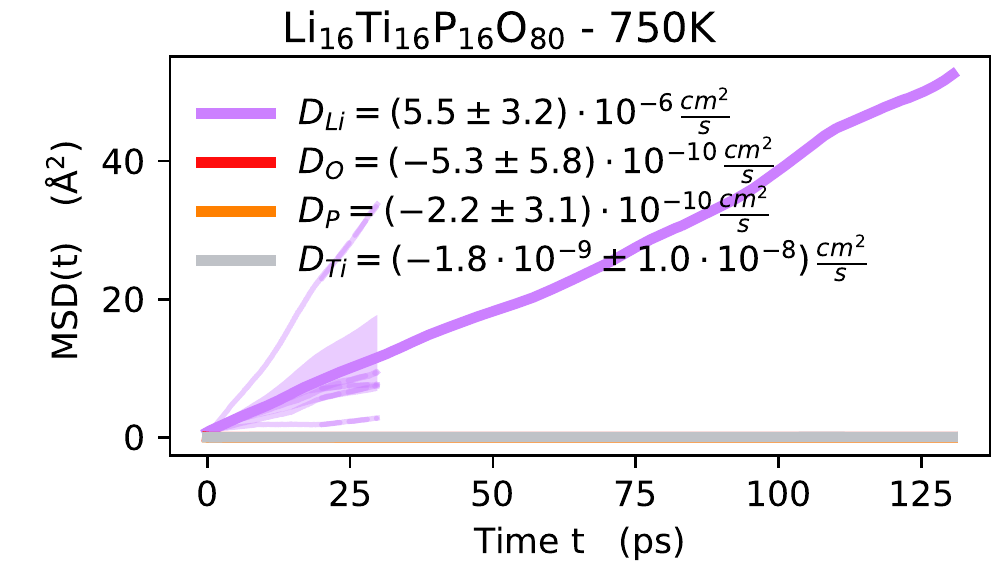}
    \caption{MSD of all species in \ce{Li16Re24S44} (left) and \ce{Li16Ti16P16O80} (right) at 750~K from FPMD.}
    \label{fig-msd-B1}
\end{figure}

\begin{figure}[t]
    \centering
    \includegraphics[width=0.35\hsize]{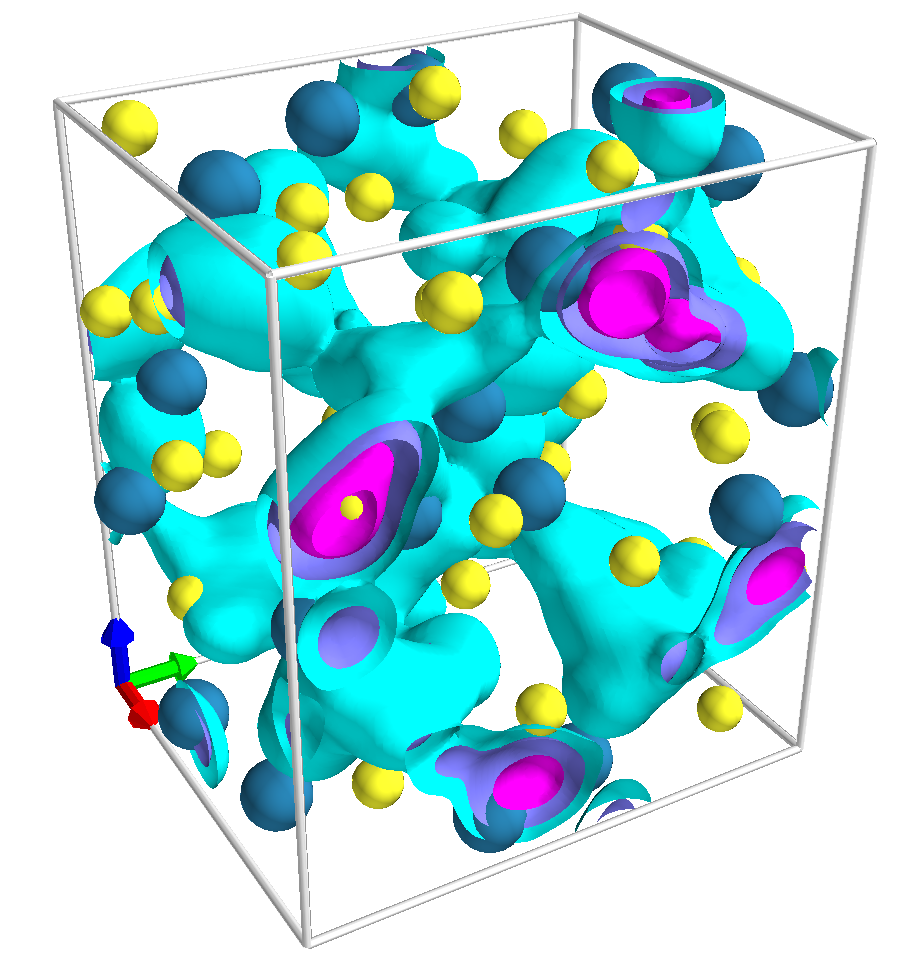}
    \includegraphics[width=0.63\hsize]{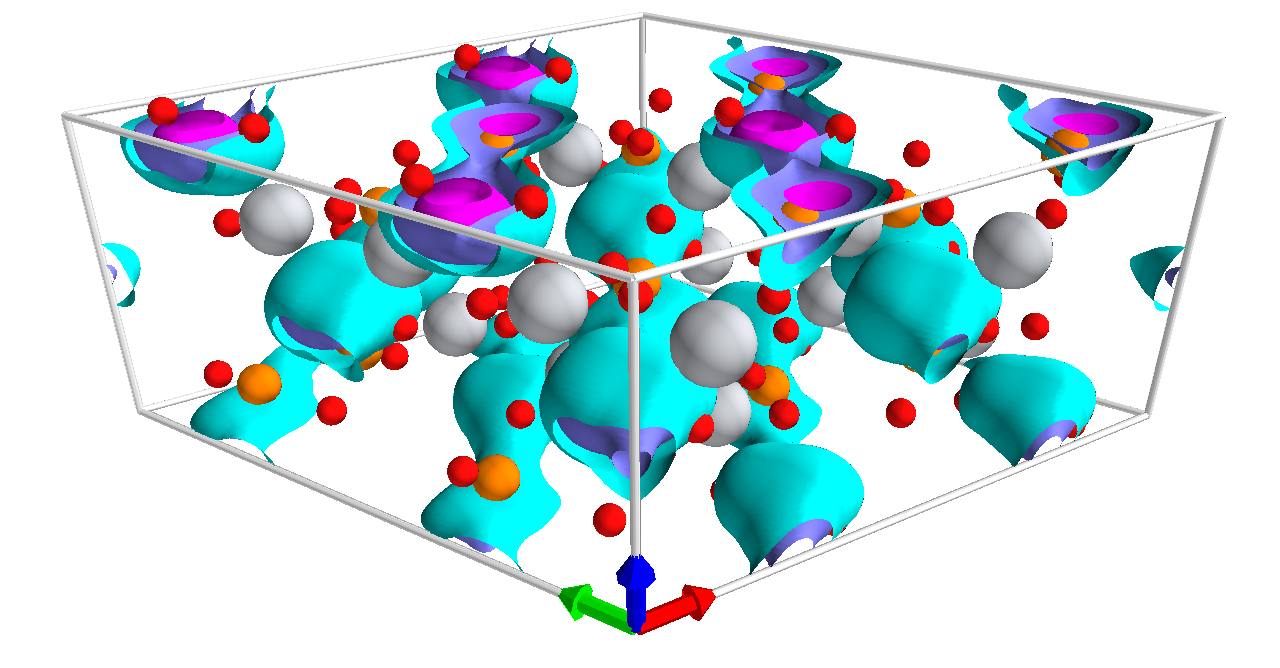}
    \caption{Li-ionic density of \ce{Li16Re24S44} (left) and \ce{Li16Ti16P16O80}  (right) at 500~K from FPMD.}
    \label{fig-dens-Li16Re24S44-500}
\end{figure}

We portray in the following all materials where we observe ionic diffusion at high temperature, but where our FPMD simulations either do not show significant diffusion at low temperature, or we cannot resolve the low-temperature diffusion, either because we cannot reach the necessary time scales with FPMD, or  we did not attempt to simulate the low-temperature regime. 
We order the results by the likelihood that these materials show fast-ionic diffusion at all temperatures.
We stress that many of the materials might show significant ionic conductivity in experiments also at low temperatures.
The inability of our simulations to resolve the diffusion at lower temperature is due to the fact that structures with a lower diffusion coefficient require longer simulation times, that cannot be afforded with FPMD.
For example, Murugan and Weppner~\cite{murugan_fast_2007} report an ionic conductivity of 0.02~S\,cm$^{-1}$ for cubic LLZO at 500~K, which results in a tracer diffusion $2\times 10^{-7}$~cm$^2$\,s$^{-1}$, assuming a Haven ratio of 1~\cite{marcolongo_ionic_2017}.
Given this diffusion coefficient, an ion travels in 100~ps on average a mean-squared distance of $6 D_{tr} t \approx 1$~\r{A}$^2$.
It is obvious that a simulation of $\sim$100~ps cannot resolve the diffusive behavior of Li ions in the garnet at 500~K.
Therefore the following materials could be 
ionic conductors also at ambient temperatures.

\paragraph*{\ce{Li4Re6S11}:} We study \ce{Li4Re6S11}, first synthesized by Bronger \textit{et al.}~\cite{bronger_darstellung_1985} in its supercell \ce{Li16Re24S44}. 
We observed a volume expansion of 2.61\% during the cell relaxation. 
Our simulations at 500~K (for 87.1~ps), 600~K (174~ps), 750~K  (87.2~ps) and 1000~K (290.8~ps) give evidence for high ionic diffusion. We show this at 750~K in \myfigref{fig-msd-B1}, and for the other temperatures in \appfigref{fig-msd-Li16Re24S44}.
The short simulations do not allow to resolve the diffusion at low temperature, which forbids us to classify this material into the group fast-ionic conductors. 
Nevertheless, the molecular dynamics dynamics are compatible with diffusive behavior also at low temperature.
In addition, we find the host lattice to be  stable during the dynamics.
We show the Li-ion density, calculated from the simulation at 500~K, in the left panel of \myfigref{fig-dens-Li16Re24S44-500}, which classifies this material as a three-dimensional ion conductor.

\paragraph*{\ce{LiTiPO5}:} The  oxyorthophosphate  $\alpha$-\ce{LiTiPO5}~\cite{gejfman_crystal_1993} is studied in its supercell \ce{Li16Ti16P16O80}. 
We observed a volume expansion of 5.76\% during the cell relaxation. 
The host lattice is dynamically stable and shows no diffusion.
Our simulations of at least 232.4~ps show highly diffusive behavior for the Li ions at high temperature, and significant diffusion diffusion at low temperature, but are not accurate enough to allow for a quantitative results at low temperature.
The MSD obtained from our simulations is shown in the right panel of \myfigref{fig-msd-B1} for 750~K (other temperatures in \appfigref{fig-msd-Li16Ti16P16O80}).
The Li-ion density at 500~K, illustrated in the right panel of \myfigref{fig-dens-Li16Re24S44-500}, gives evidence for uni-dimensional conductive pathways in this material.

\paragraph*{\ce{Li6PS5I}:} The lithium argyrodite
\ce{Li6PS5I} is simulated in the supercell of \ce{Li48P8S40I8}.
We are not able to resolve the diffusion in the diffusive regime precisely, and classify also this material as a potential ionic conductor.
We became aware afterwards that its ionic conductivity is known from experiments and simulation~\cite{oliver_pecher_atomistic_2010}.

\begin{wrapfigure}{r}{0.38\hsize} 
    \includegraphics[width=\hsize, clip, trim={0cm 4cm 0cm 2cm}]{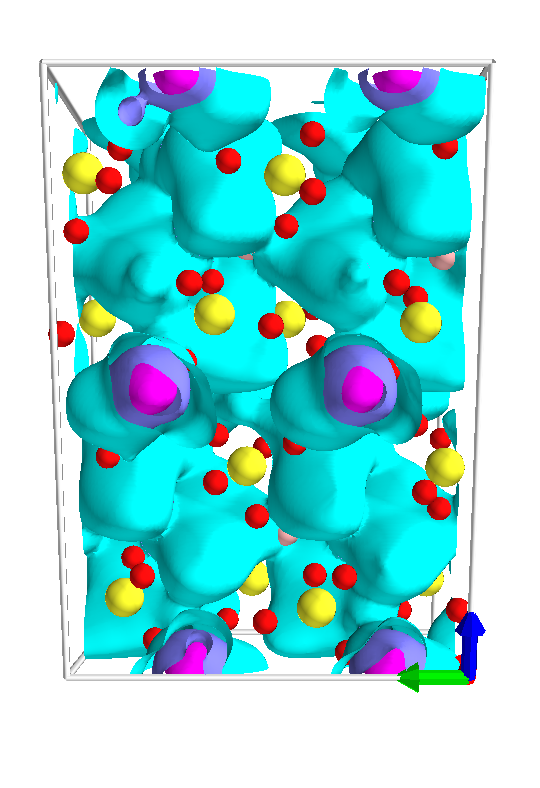}
    \caption{Li-ionic density of \ce{Li20B4S16O64} at 600~K.}
    \label{fig-dens-Li20B4S16O64-600}
\end{wrapfigure} 

\paragraph*{Others:} We report in the following the materials that show significant diffusion at high (1000~K) and intermediate (600~K and 750~K) temperatures, but no diffusion at low (500~K) temperature.
(1) The first is \ce{Li5BS4O16}~\cite{daub_michael_further_2014}, studied in its supercell \ce{Li20B4S16O64}.
Our simulations of 218~ps at 1000~K and 610~ps at the lower temperatures give evidence (see \appfigref{fig-msd-Li20B4S16O64}) for fast-ionic diffusion at high and intermediate temperatures, within a stable host lattice. 
The diffusion is three-dimensional, as illustrated by the Li-ion density at 600~K, given in \myfigref{fig-dens-Li20B4S16O64-600}.
At 500~K, there is no or negligible diffusion observed in the simulation.
(2) We simulate \ce{LiTaGeO5}~\cite{malcherek_structure_2002} in the supercell \ce{Li4Ta4Ge4O20} for at least 145~ps at the four different temperatures.
The reference reports a phase transition at 231~K to a disordered phase. Our conclusions only adhere to the ordered phase.
The structure shows high ionic diffusion at high and intermediate temperatures, but our 500~K simulation show no diffusive behavior.
The respective MSDs computed from the trajectories are shown in \appfigref{fig-msd-Li4Ta4Ge4O20}. 
(3) For \ce{LiIO3}, which we study in its supercell \ce{Li8I8O24}, we observe a large volume expansion of 15\%, and significant host-lattice diffusion.
The diffusion of Li ions is considerable at high temperatures, but drops to negligible diffusion at low temperature (see \appfigref{fig-msd-Li16I16O48}).

\begin{wrapfigure}{R}{0.55\hsize} 
    \includegraphics[width=0.49\hsize]{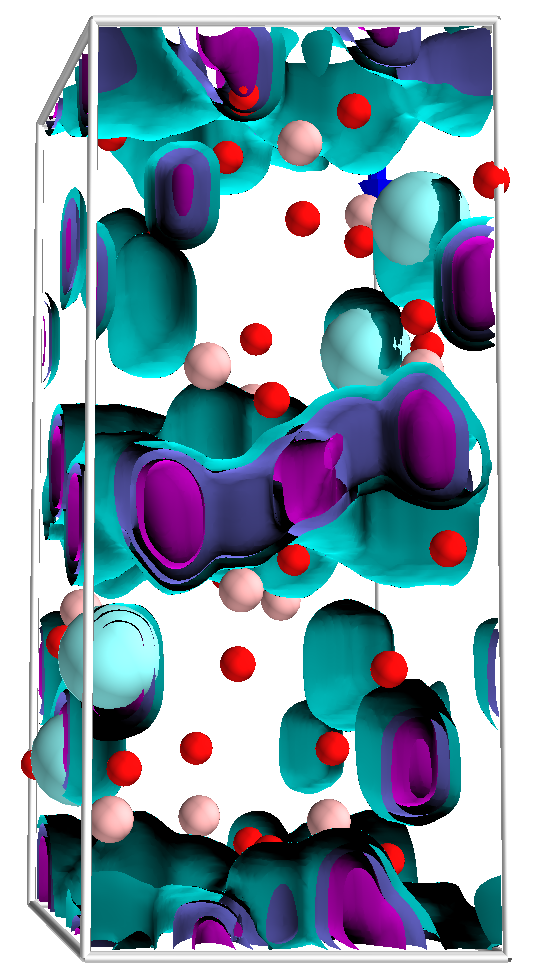}
    \includegraphics[width=0.49\hsize]{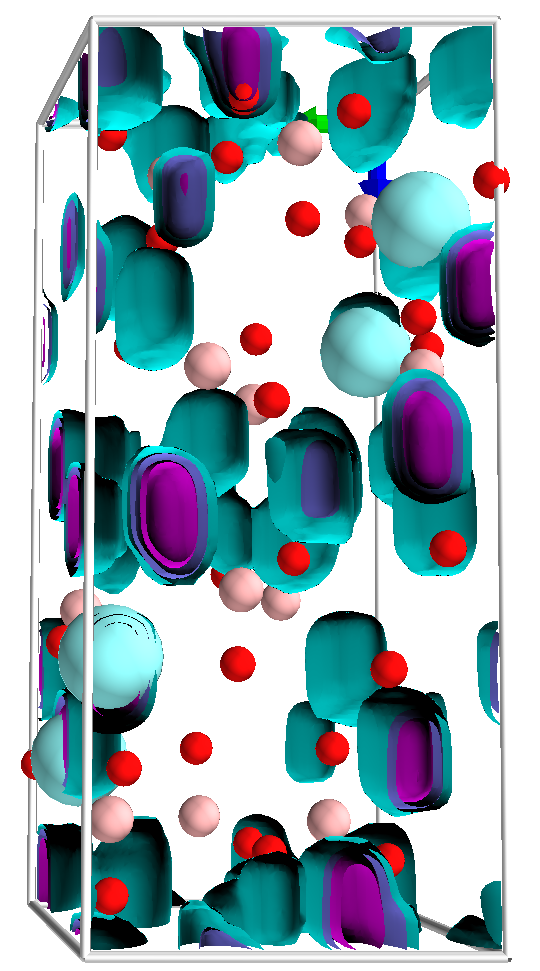}
    \caption{Li-ionic density of \ce{Li24Y4B12O36} at 750~K in the left panel and 500~K on the right panel.}
    \label{fig-dens-Li24Y4B12O36}
\end{wrapfigure} 

\begin{wrapfigure}{R}{0.6\hsize} 
    \includegraphics[width=\hsize]{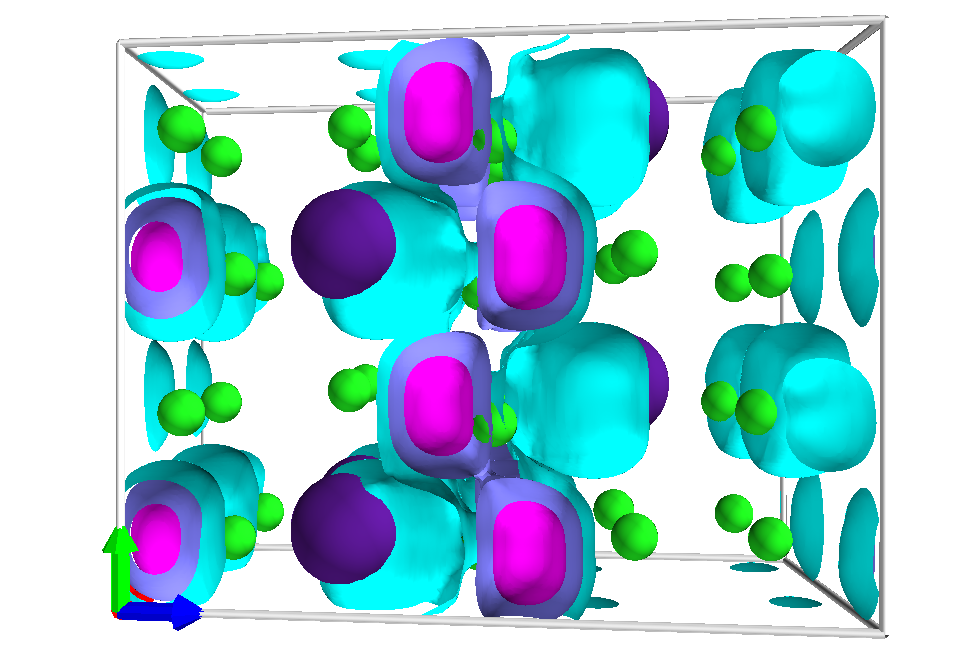}
    \caption{Li-ionic density of \ce{Li24Cs8Cl32} at 750~K.}
    \label{fig-dens-Li24Cs8Cl32-750}
\end{wrapfigure}

In the following, we report the candidate materials that show significant diffusion at high temperature, but where the diffusion becomes very small or negligible at intermediate temperatures (600~K or 750~K).
(1) The $\mathrm{\beta}$-eucryptite \ce{LiAlSiO4} was first referenced by Pillars \textit{et al.}~\cite{william_w._pillars_crystal_1973}, and its  ionic conductivity was studied experimentally in 1980~\cite{press_neutron_1980}, where a uni-dimensional transport mechanism was observed.
A more recent computational and experimental study~\cite{singh_superionic_2017} confirmed this behavior.
We also observe the uni-dimensional transport along the c-axis of this material with our simulation in the supercell \ce{Li12Al12Si12O48}. 
The host-lattice remains stable during the dynamics, whereas Li ions show significant diffusion at high temperature. At 600~K the diffusion is negligible (see \appfigref{fig-msd-Li12Al12Si12O48}). 
(2) \ce{Li2Mg2S3O12}~\cite{touboul_structure_1988}, that we studied in its supercell \ce{Li8Mg8S12O48} also shows significant diffusion at high temperature, that however drops significantly when lowering the temperature, and becomes very low around 600~K.
A derived structure (doped with Fe and V) was studied as a cathode material~\cite{morgan_experimental_2002}.
(3) We study the perthioborate~\cite{jansen_na2b2s5_1995} \ce{Li2B2S5} in its supercell, \ce{Li8B8S20}.
In this structure, Li-ions are intercalated between layers of \ce{Be2S5}$^{2-}$.
The diffusion at high temperature is substantial (see \appfigref{fig-msd-Li8B8S20}).
Despite the low diffusion at low temperatures, promising results at high temperature for \ce{Li2B2S5} suggest that the family of perthioborates could be studied more extensively.
(4) \ce{Li6Y(BO3)3} was researched for potential application in solid-state lasers~\cite{luo_study_1991}.
We simulate its supercell \ce{Li24Y4B12O36} with FPMD, and observe fast-ionic diffusion at 1000~K and 750~K, which is shown in \appfigref{fig-msd-Li8B8S20}.
At lower temperature (600~K and below), the diffusion becomes too small detect with FPMD.
From the Li-ion densities (shown in \myfigref{fig-dens-Li24Y4B12O36}) it is evident that in-plane diffusive pathway in this material is no longer active at lower temperatures.
(5) We simulate \ce{Li3CsCl4}~\cite{pentin_theoretical_2012} in its supercell, \ce{Li24Cs8Cl32}.
The two-dimensional ionic diffusion at elevated temperatures (see \myfigref{fig-dens-Li24Cs8Cl32-750}) drops substantially at 600~K, as shown in \appfigref{fig-msd-Li24Cs8Cl32}.
(6) Also the ortho-diphosphate \ce{Li9Ga3(P2O7)3(PO4)2}~\cite{liu_layered_2006}, studied as \ce{Li18Ga6P16O58}, displays significant diffusion at high temperatures, but negligible diffusion at 600~K (see \appfigref{fig-msd-Li18Ga6P16O58}).
A Va-analogue of this structure was researched~\cite{balasubramanian_facile_2017} as a cathode material.

In the following, we describe structures that show only diffusion at high temperature (1000~K), but no diffusion at lower temperatures (750~K, 600~K, and 500~K).
(1) As a first example, \ce{Li7RbSi2O8} was synthetized accidentally by Bernet and Hoppe~\cite{bernet_lithosilicat_1991} as a new type of orthosilicate. We simulate the supercell \ce{Rb8Li12B4P16O56} and observe diffusion only at high temperature.
(2) The lithium disulfate \ce{Li2S2O7}~\cite{logemann_crystal_2014}, studied as \ce{Li16S16O56}, also shows diffusion only at 1000~K and none at the lower temperatures.
(3) \ce{Li2In2GeS6} and (4) \ce{Li2ZnSnSe4} were researched independently~\cite{yin_synthesis_2012,zhang_infrared_2015} for the nonlinear optical properties in the infrared spectrum.
We simulated the respective supercells \ce{Li16In16Ge8S48} and \ce{Li16Zn8Sn8Se32}, and both structures show diffusion at 1000~K and no diffusion at lower temperatures, as can be seen in \supplref{fig-msd-Li16In16Ge8S48} and 
\supplref{fig-msd-Li16Zn8Sn8Se32}.
Also (6) \ce{Li3GaF6}, studied as \ce{Li18Ga6F36}, (7) \ce{Li2T3O7}, studied as (8) \ce{Li8Ti12O28}, and (9) \ce{LiMoAsO6}, studied as \ce{Li8Mo8As8O48} show diffusion only at high temperature. The MSD are shown in Figs.~\ref{fig-msd-Li18Ga6F36},~\ref{fig-msd-Li8Ti12O28}, and~\supplref[]{fig-msd-Li8Mo8As8O48}, respectively.

We also study an additional 21 structures at 1000~K alone, without simulating the diffusion also at lower temperatures.
We find them to show non-negligible diffusion, but the diffusion is not high enough to warrant a costly estimate of the diffusion at lower temperature and extraction of the barrier, or because very similar compounds have their activation barrier computed.
We find diffusion in the doped halides \ce{LiGaCl3}~\cite{honle_preparation_1988}, \ce{LiGaBr4}~\cite{honle_darstellung_2014},  \ce{Li6MgBr8}~\cite{schneider_kristallstrukturen_2014}, and \ce{LiAuF4}~\cite{engelmann_darstellung_1991}.
We also calculate the diffusion of Li ions in the phosphides \ce{Li3P7}~\cite{manriquez_trilithiumheptaphosphid_1986} and \ce{LiP7}~\cite{v._schnering_lithiumphosphide_1972},
the sulfides \ce{Li3AsS3}~\cite{seung_li3ass3_1998} and \ce{LiRb2TaS4}~\cite{huang_syntheses_2005}, and the selenides \ce{LiAlSe2}~\cite{kim_synthesis_2000} and \ce{Li2In2SiSe6}~\cite{yin_synthesis_2012}.
The same applies to the borate \ce{LiBS4Cl4O12}~\cite{mairesse_lithium_1980} and the NASICON-type phosphate \ce{LiSn2P3O12}~\cite{morin_phase_1998},
the germanates \ce{Li4G9O20}~\cite{vollenkle_crystal_1971} \ce{Li2Ge4O9}~\cite{redhammer_polar_2013},
the 
metaperiodate \ce{LiIO4}~\cite{kraft_kristallstruktur_1995},
the 
phosphates \ce{Li4P207}~\cite{daidouh_new_1997} and \ce{LiInP2O7}~\cite{tran_qui_synthese_1987},
the
phenakyte \ce{Li2SeO4}~\cite{hartman_uniform_2015},
the
titanate \ce{Li4TiO4}~\cite{gunawardane_crystal_1994},
the silicate
\ce{Li6Si2O7}~\cite{vollenkle_kristallstruktur_1969},
and the
borosulfonate \ce{LiBS2O8}~\cite{daub_exploring_2013}.
 The materials are listed, with the supercell employed, the originating database entry, the volume expansion, and the simulation length at 1000~K in \suppltabref{tab-groups-B-only-1000}.

\subsection{Non-diffusive structures}
We also find 70 materials to be not diffusive at 1000~K in our FPMD simulations.
These are structures where our simulations give evidence that this structure will also not conduct in experiment, unless doped significantly.
The materials are listed in Table~\ref{tab-groups-Ca-only-1000} of the SI, together with the respective ICSD/COD entry, the volume change, and the simulation time. 
In the following, we only give a brief summary of the materials.

We observe no diffusion in the doped nitrides
\ce{Li2CeN2},
\ce{Li5ReN4}~\cite{chaushli_li5ren4_2000},
\ce{Li7PN4},
\ce{Li3ScN2}~\cite{rainer_niewa_li3[scn2]:_2003},
\ce{Li7NbN4}~\cite{vennos_structure_1992},
\ce{Li3AlN2}~\cite{juza_ternaren_1948},
\ce{Li6WN4}~\cite{yuan_synthesis_2005}, or
\ce{Li4TaN3}~\cite{niewa_preparation_2002}, nor in the niobium-doped oxynitride \ce{Li16Nb2N8O1}~\cite{cabana_exploring_2010}.
The doped halides \ce{LiAuF4}~\cite{hoppe_neue_1970} and \ce{LiInNb3Cl9}~\cite{lachgar_synthesis_1994} also show no diffusive behavior.
The borates
\ce{Li2AlB5O10}~\cite{he_li2alb5o10_2001},
\ce{Li3Sc(BO3)2}~\cite{mao_trilithium_2008},
\ce{Li3GaB2O6}~\cite{smith_crystal_2017},
\ce{Li2AlBO4}~\cite{psycharis_crystal_1999},
\ce{Li3GaB2O6}, and \ce{Li8Be5B6O18}~\cite{wang_three_2014} are poorly diffusing in FPMD simulations, as is also the 
borophosphate \ce{Li2NaBP2O8}~\cite{hasegawa_synthesis_2015}.
We also consider the phosphates
\ce{Li2Cd(PO3)4}~\cite{averbuch-pouchot_structures_1976},
\ce{LiPO3}~\cite{ben-chaabane_li6p6o18:_1998}, \\
\ce{Li4ZnP2O8}~\cite{jensen_hydrothermal_2002}, and
\ce{Li9Mg3F3P4O16}~\cite{yahia_crystal_2014}, as poor ionic conductors at full Li-ion occupation. For the last case, this is in contradiction to the reported ionic conductivity from experiment~\cite{yahia_crystal_2014}. 
The phosphide 
\ce{Li4SrP2}~\cite{dong_synthesis_2007} is a poor ionic conductor, based on our simulations.
The same applies to the silicates
\ce{LiBSi2O6}~\cite{parise_hydrothermal_2000},
\ce{L2Si3O7}~\cite{kruger_li2si3o7:_2007},
\ce{LiYSiO4},
\ce{Li3AlSiO5}~\cite{chen_li_2016}, and 
\ce{Li2Si2O5}~\cite{de_jong_mixed_1998}, 
and the tellurates
\ce{Li3TeO3}~\cite{folger_kristallstruktur_1975},
\ce{Li4TeO5}~\cite{untenecker_neues_1987-1}, and
\ce{Li6TeO6}~\cite{wisser_neues_1989},
Also the tantalate
\ce{Li6Sr3Ta2O11}~\cite{bharathy_crystal_2008}, studied for its photoluminescence, is a poor ionic conductor, as is the La-doped carbonate
\ce{LiLaC2O3}~\cite{glaser_improved_2010},
or the K-doped aluminate
\ce{Li4KAlO4}~\cite{grebe_oxid_1988}.
The zincates, molybdates and arsenates we classify as non-diffusive are:
\ce{Li2MoO4}~\cite{s.yip_spontaneous_2010},
\ce{LiZnAsO4}~\cite{jensen_hydrothermal_1998},
\ce{Li3AlMo2As2O14}~\cite{hajji_li3almoo22o2aso42_2009},
\ce{LiKZnO2}~\cite{baier_zur_1985},
\ce{Li6ZnO4}~\cite{untenecker_neues_1987}. The last structure was confirmed to not be an ionic conductor unless doped with Nb~\cite{konovalova_new_2009}.
Additional oxides that are poor ionic conductors are the germanate
\ce{Li4Ge5O12}~\cite{greenberg_structure_1990} and  the phenakite \ce{Li2WO4}~\cite{hasegawa_synthesis_2015}.
We also find
\ce{Li3AuO3}~\cite{wasel-nielen_zur_1970} to not be diffusive. Interestingly, a study by Fils\o \textit{et al.}~\cite{filso_visualizing_2013} found this materials to be a threedimensional ionic conductor using an approximate electron-density descriptor.
At full occupation with lithium, we do not see any diffusion.
This is also the case for the niobate \ce{Li4KNbO5}~\cite{wehrum_erste_1993}, as well as 
\ce{Li2PdO2}~\cite{wolf_notiz_1986} and \ce{Li8PtO6}~\cite{kroeschell_neue_1986}.

\subsection{Structures diffusive in the pinball model}

We have 14 structure where we can not make a statement from FPMD, mostly due to frequent failures during the dynamics, especially during the self-consistent minimization of the electronic charge density.
The materials are given in \suppltabref{tab-groups-D-only-1000} and summarized in the following.

Materials that show ionic diffusion in the pinball model are the molybdate \ce{Li4Mo3O8},
the silicate  \ce{LiTaSiO5}~\cite{genkina_crystal_1992},
the ortho-diphosphate,\ce{Li2P2PdO7}
the nalipoite \ce{Li2NaPO4}
the layered borate \ce{Li3BaNaB6O12}~\cite{chen_synthesis_2012},
the borates \ce{LiNaB4O7}~\cite{maczka_crystal_2007}
and \ce{Li2NaBO3}~\cite{miessen_neue_1987}, and the boracite-type \\
\ce{Li10B14Cl2O25}~\cite{vlasse_crystal_1981}, where a transport mechanism is observed in the experiment.
We also cannot study he doped halides
\ce{LiAuI4}~\cite{lang_synthese_1997},
\ce{Li3ScF6}~\cite{tyagi_syntheses_2005}, and
\ce{LiNb3Cl8}~\cite{bajan_two-dimensional_1997} with FPMD.
The same applies to \ce{LiAuS4O14}~\cite{logemann_unique_2011},
the NASICON ~\ce{LiZr2As3O12}~\cite{petkov_synthesis_2014}, and
\ce{LiAlGeO5}~\cite{tripathi_synthesis_2000}.
For the latter, a study employing CPMD shows no Li-ion diffusion~\cite{ceriani_role_2003}.

\subsection{Context of this screening}

As described in \mysecref{sec-methods}, we screen all the Li-containing materials in the ICSD and COD experimental structure databases without partial occupancies on any site and at full Li stoichiometry for solid-state ionic conductors.
This work is therefore not a complete screening of these databases. For example, another 645 (not necessarily unique) structures in ICSD \& COD show partial occupancy only on Li sites and full occupancy on sites occupied by other species.

A structure with full occupancy on all Li sites should not be ionically conducting if we neglect Li diffusion via interstitial sites.
Only upon the introduction of Li vacancies could the  vacancy-assisted Li-ionic diffusion be unlocked.
However, the fact that we find several ionic conductors can be explained.
First, we conduct our computational experiments at elevated temperature, for which the Li occupations reported in CIF files, usually measured at room temperature, are merely indicative.
Second, due to inherent difficulties in determining the Li-ion positions in XRD, many low-occupancy sites might have been discarded when reporting the structure.
Therefore, the class of Li-ionic system at full stoichiometry and without partial occupancies contains Li-ionic conductors, as we show.
Of course, screening for solid-state ionic conductors within another class of materials (systems reported with partial occupancies) should result in more candidates for solid-state electrolytes.
Furthermore, the creation of Li vacancies in all materials should result in new candidates, since a high diffusion upon the introduction of vacancies indicates that the structure could be doped into high ionic conductivity in experiment.
This can be done for structures with or without partial occupancies.
In short, the tools and methodology presented here can be used to study materials with partial occupancy at varying Li-ion concentration, for a complete screening of structural repositories for novel solid-state ionic conductors.

\begin{wrapfigure}{r}{0.55\hsize} 
\includegraphics[width=\hsize]{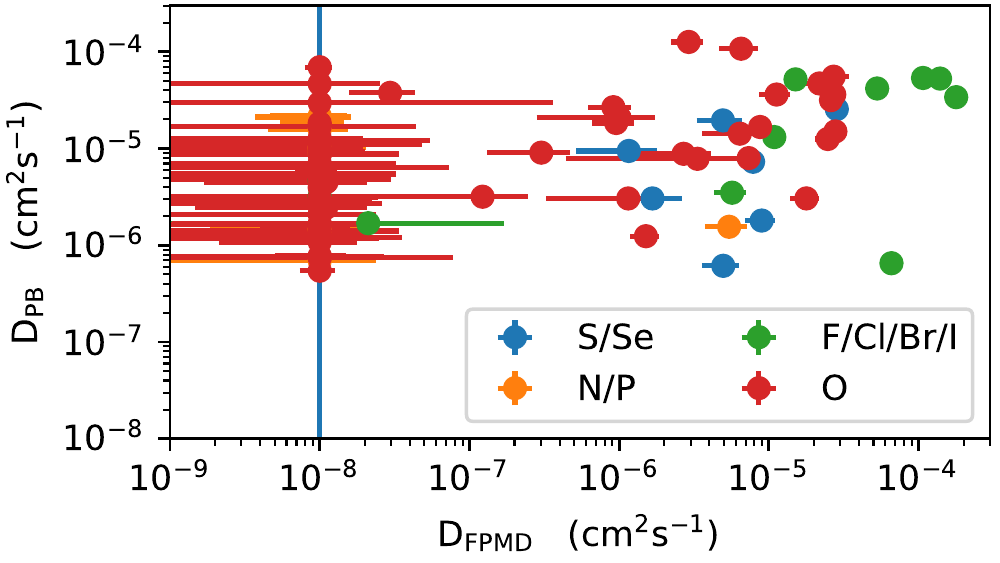}
\caption{
Diffusion FPMD vs pinball model used in the screening at 1000~K for 95 materials, color-coded by the predominant anion.
The vertical line at $10^{-8}$ gives the limit below which we cannot converge the slope of the MSD with FPMD, therefore we set this as the lower boundary for diffusion.}
\label{fig-diff-fpmd-pb-loc-rand}
\end{wrapfigure}

This first screening also allows us also to estimate how well the pinball performs in a screening scenario, since we have plenty of data to compare FPMD and pinball simulations for the same structures.
In \myfigref{fig-diff-fpmd-pb-loc-rand} we show the diffusion at 1000~K for 95 structures computed with FPMD ($\mathrm{D_{FPMD}}$ on $x$-axis) and the pinball model ($\mathrm{D_{PB}}$ on $y$-axis). 
The diffusion coefficient from FPMD is not well reproduced by the pinball model, which can have several causes.
First, the local pinball could be inaccurate, compared to the non-local pinball~\cite{kahle_modeling_2018}.
Second, fitting the pinball model with snapshots from random displacements could result in parameters $\alpha_{1,2}$ and $\beta_1$ that do not reproduce the forces on Li ions accurately during the dynamics.
Third, freezing the charge density could have a larger effect on the Li-ion dynamics than in the prototypical ionic conductors studied beforehand~\cite{kahle_modeling_2018}.
Last, freezing the host lattice could also have a larger effect on the Li-ion dynamics than anticipated.
For another screening with the pinball model, especially at varying stoichiometry, the accuracy of the pinball model should be improved by studying these effects individually.
Nevertheless, while we see many false positives in our screening, it is critical to state that so far no false negatives have emerged.
A $\sim$50\% ratio of false positives is very much tolerable for screening application, and outweighed by the low computational cost of the pinball, for which we give additional details in \appsecref{app-comput-cost}.

\section{Conclusions \& outlook}
\label{sec-conclusions}


We presented a computational screening of two large repositories of experimental structures, the ICSD~\cite{belsky_new_2002} and COD~\cite{grazulis_crystallography_2012}, totalling  $\sim$1'400 unique crystal structures.
We used the Kohn-Sham band gap from density-functional theory at the PBE level to find electronically insulating systems and filtered for systems that are likely to display fast-ionic diffusion by harnessing the computational efficiency of the pinball model~\cite{kahle_modeling_2018} with molecular dynamics,
which captures the collective effects of Li-ion migration, totalling 7.6~$\mu$s of simulation time.
About 130 structures that showed high Li-ion diffusion in the pinball model were simulated with accurate first-principles molecular dynamics for a total of 45~ns at high and intermediate temperatures, enabling also the extraction of the activation energy from first principles.
We found five materials with fast ionic diffusion, some in the range of the well-known superionic conductor \ce{Li10GeP2S12}, as for example the Li-oxide chloride \ce{Li5Cl3O}, the doped halides \ce{Li2CsI3}, \ce{LiGaI4}, and \ce{LiGaBr3}, or the Li-tantalate \ce{Li7TaO6}. 
We also found 40 materials that show significant diffusion at 1000~K, but where we cannot rigorously extract the barrier due to the short time scales accessible to FPMD, such as \ce{Li4Re6S11} and \ce{LiTiPO5}.
These potential fast-ionic conductors could be studied further, in more detail, by experiments and simulations, and could result in new fast-ionic conductors or even electrolytes for next-generation solid-state Li-ion batteries.



\section*{Acknowledgements}
We gratefully acknowledge support from the Swiss National Science Foundation (MARVEL NCCR and project 200021-159198) and the Swiss National Supercomputing Centre CSCS (project s836).

\newpage
\section{Method details}

\subsection{Duplicate filter parameters}
\label{app-meth-duplicates}

All structures that have the same stoichiometric formula are compared.
Structures are marked as equal if one of the structures can be mapped into the other by the pymatgen structure matcher~\cite{ong_python_2013}, using an angle tolerance of $5^\circ$, a relative lattice tolerance of 20\%, and a site tolerance of 30\%.
For all structures found to be equal, one representative is chosen randomly, while all other structures are marked as duplicates and are not processed further in the screening.

\subsection{Composition filters}
\label{app-meth-composition-filters}
Additional filters are applied to remove structures that do not meet the the following criteria:
First, we include only structures that contain the anions N, O, F, P, S, Cl, Se, Br, and I.
Second, we only keep structures that do not contain hydrogen, since the pinball model has never been tested for hydrogen-containing Li-ionic conductors.
Physical intuitions suggests that hydrogen -- being lighter than lithium -- should yield to lithium motion, which is not compatible with he frozen framework of the model.
Third, we remove structures that contain noble gas atoms (He, Ne, Ar, Kr, Xe, or Rn), 3d-transition metals (V, Cr, Mn, Fe, Co, Ni, or Cu) due to their capability to change oxidation states, or elements that are radioactive (Tc, Po, Rn, Ac, Th, Pa, U).
Fourth, we remove structures that have elements above mercury in the periodic table, since the applications as SSE would be very unlikely.
Last, we apply a filter checking whether there are enough anions that can accept the valence electron of lithium.
For each structure, we add the number of Li ions to the number of anions multiplied by their most common oxidation states ($-1$ for halogens, $-2$ for chalcogens, $-3$ for pnictogens).
If the final number is above 0, we reject the structure because we assume not all Li atoms will not find electron acceptors for charge transfer.

\subsection{Bond distance filters}
\label{app-meth-bond-filters}

For every structure, we calculate all bond distances $\{\mathrm{A-B}\}$ between species A and B, so we can select only compositions that show bond distances that are compatible with inorganic materials.
We remove structures with $\mathrm{C-N < 1.6}$~\r{A} (cyanide group); $\mathrm{F-F} < 1.5$~\r{A}, $\mathrm{Cl-Cl} < 2.1$~\r{A}, $\mathrm{Br-Br} < 1.6$~\r{A},  $\mathrm{I-I}< 2.8$~\r{A} (halide molecules); $\mathrm{C-C} < 1.6$~\r{A} (carbon double/triple bonds); $\mathrm{O-O} < 1.6$~\r{A} (peroxide group); $\mathrm{X_i-X_j} < 0.8$~\r{A}, where $\mathrm{X_{i/j}}$ can be any element, to remove structures with any bonds being shorter than the $\mathrm{H-H}$ bond.

\subsection{Electronic structure}
\label{app-meth-electronic}
The Marzari-Vanderbilt cold smearing~\cite{marzari_thermal_1999} is set for all calculations to $\sigma=0.02$~Ry~$ \approx 0.27$~eV, and we augment the number of valence bands by 20\%.
The Brillouin-zone is sampled with a Monkhorst-Pack grid with a density of 0.2~\r{A}$^{-1}$.
If the lowest-energy state above the Fermi-level is occupied by more than $10^{-3}$ of an electron we classify that structure as not electronically insulating and reject it from the candidates.

\subsection{Variable-cell relaxation}
\label{app-meth-vc}
After applying an initial random distortion, with distortions taken from the normal distribution with $\sigma=0.1\AA$ to break crystal symmetries,
we apply the BFGS algorithm as implemented in Quantum ESPRESSO to converge the crystal structure until all following criteria have been met.
First, the forces on the atoms need to be converged to below $5\times 10^{-5}$~Ry\,bohr$^{-1}$;
Second, the total energy difference between consecutive iterations needs to be below $1\times 10^{-4}$~Ry;
Third, the pressure has to be less than 0.5~kbar.
Kpoint-grids are chosen as explained in \mysecref{app-meth-electronic}.
No valence bands are added, and no smearing is applied, since only electronic insulators are relaxed.
Van-der-Waals contributions are not considered.

\subsection{Supercell creation}
\label{app-supercells}
All possible supercells are built by expanding the unit cell vectors of the primitive cell $\bm a_{p_1}$,  $\bm a_{p_2}$ and  $\bm a_{p_3}$
to the supercell vectors $\bm a_{s_1}$,  $\bm a_{s_2}$ and  $\bm a_{s_3}$ via an expansion by a $3\times 3$ matrix $\bar{R}$ of integers~\cite{hart_algorithm_2008}:
\begin{align}
    	\begin{pmatrix}
       \bm a_{s_1},
       \bm a_{s_2}, 
       \bm a_{s_3}
    \end{pmatrix} =
    \begin{pmatrix}
       R_{11} & R_{12} & R_{13} \\
       R_{21} & R_{22} & R_{23} \\
       R_{31} & R_{32} & R_{33} \\
    \end{pmatrix} 
    \begin{pmatrix}
       \bm a_{p_1}, 
       \bm a_{p_2}, 
       \bm a_{p_3}
    \end{pmatrix}. 
\label{eq-supercell}
\end{align}
We apply the criterion that the supercell needs to enclose a sphere of a diameter $d_{inner}$ (distance criterion) that defines a minimal distance of interaction of a particle with periodic images.
We find the coefficients $R_{ij} \in \mathbb{Z}$ of $\bar{R}$ that minimize the volume of the cell under this constraint, using our implementation of \textsc{supercellor}~\cite{kahle_supercell_2019}.

\subsection{Fitting}
\label{app-fitting-pinball}

After preliminary tests, it was determined that 5000 force components are needed for accurate fitting.
As an example, $\mathrm{Li_{20}Ge_2P_4S_{24}}$ has 20 Li ions, which results in $\lceil 5000/(3\cdot 20) \rceil = 84$ uncorrelated snapshots being needed for the fitting ($y = \lceil x \rceil$ refers to the ceiling function: the output $y\in\mathbb{Z}$ of this function is the smallest integer larger or equal to its input $x\in \mathbb{R}$).
Since the generation of these snapshots with BOMD defeats the purpose of an efficient screening, we create the snapshots by displacements of the Li ions from equilibrium taken from a normal distribution with $\sigma=0.1$~\r{A}.
For each such configuration, one calculation in the pinball model and one calculation with DFT is performed.
Only the forces are used to regress the parameters $\alpha_1$, $\alpha_2$, and $\beta_1$ of \myeqref{eq.pinball-screened}.

\subsection{Temperature control}
\label{app-temperature-control}
We need to control the temperature of the system we are simulating in the pinball model with minimal effects on the dynamics.
However, the number of particles that can move in pinball model is quite small, making thermalization difficult to achieve.
We also observed that systems of small Li-ion density show very slow equilibration of the energy.
Such a system resembles a system of weakly coupled harmonic oscillators, for which well-known thermostats like the  Nos\'{e}-Hoover tend to fail.
Local thermostats, such as the Andersen (stochastic collision) thermostat~\cite{hans_c._andersen_molecular_1980} can handle such cases well, but have two major disadvantages.
First, they have system-dependent parameters, such as the collision frequency, determining the strength of interaction between the modes of the system and the external bath~\cite{frenkel_understanding_1996}, 
and system-dependent parameters are problematic in any high-throughput scenario.
Second, the thermostat could suppress diffusion~\cite{tanaka_constant_1983}, depending on the collision frequency.
To prevent the dynamics being affected by a thermostat, we branch microcanonical simulations from uncorrelated snapshots of a canonical trajectory at the target temperature, a technique that has been explored for path-integral molecular dynamics~\cite{perez_comparative_2009}.

We use a timestep of $\mathrm{dt}=0.96$~fs, with snapshots being stored every 20~dt.
The collision frequency of the thermostat for the canonical simulations is set to $1000$~dt, and snapshots for the start of the microcanonical branches are taken every 3000~dt, which means that every particle's velocities are reset on average three times between consecutive snapshots.
In practice, eight snapshots are generated in one canonical run (in 24'000 timesteps) to allow for a certain degree of parallelization over microcanonical trajectories.
Therefore, eight microcanonical simulations (each for 50'000~dt) can be performed in parallel after the completion of the canonical simulation.
This operation is done at least four times (setting the minimum simulation time to $4\cdot 8\cdot 50000\cdot 0.96~\mathrm{fs}=1.5~\mathrm{ns}$), and maximally
48 times, setting the maximum simulation time to $48\cdot 8\cdot 50000\cdot 0.96~\mathrm{fs}=18.4~\mathrm{ns}$.
In between, after completion of each canonical trajectory with eight microcanonical branches,
the workflow checks  whether the error of the mean of the diffusion coefficients, estimated from all microcanonical simulations, is either converged below $1\times 10^{-8}~\mathrm{cm^2\,s^{-1}}$ or 5\% of the mean of the diffusion.
%
%

\subsection{FPMD}
\label{app-meth-fpmd}
The supercells are created from a unit cell with the distance criterion set to $d_{inner}=6.5$~\r{A} to allow for smaller cells than for the pinball simulations, due to the computational cost and scaling of FPMD.

We perform Born-Oppenheimer molecular dynamics with a timestep of $\mathrm{dt=1.45~fs}$, simulating the canonical ensemble with the stochastic velocity rescaling~\cite{bussi_canonical_2007} thermostat, implemented by us into Quantum ESPRESSO, using a characteristic decay time $\tau=100~\mathrm{dt}$ to achieve efficient thermalization.
A continuous trajectory is created using a custom AiiDA workflow that converges the error of the mean of the diffusion to below $1\times 10^{-8}~\mathrm{cm^2\,s^{-1}}$ or to below 5\% of the mean of the diffusion.
The Brillouin-zone is sampled at the $\Gamma$-point only.

\section{Computational cost of the screening}
\label{app-comput-cost}
\begin{wrapfigure}{l}{0.45\hsize}
	\centering
	\includegraphics[width=\hsize]{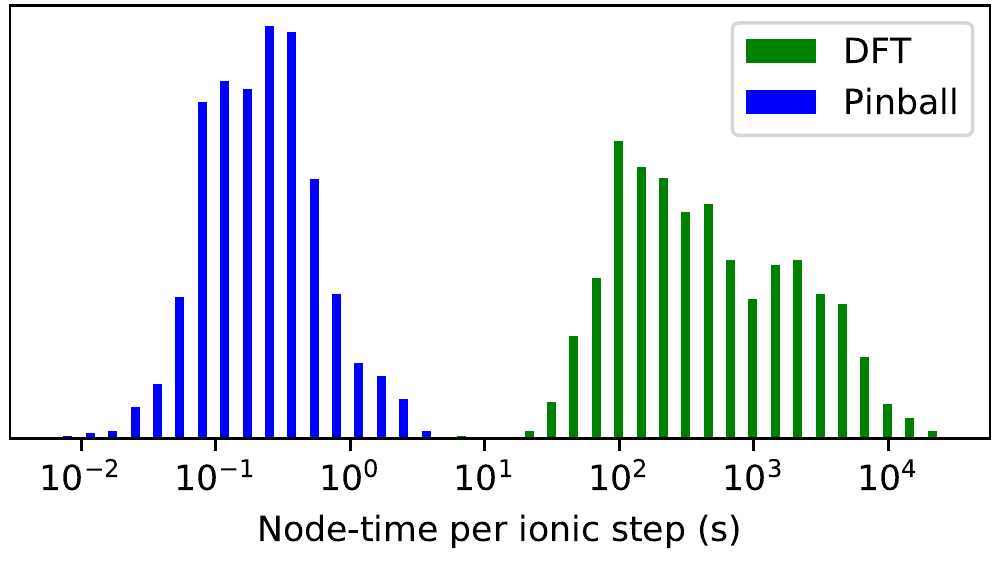}
	\caption{
	Histogram of the average node-time per ionic step for each structure, in the pinball model in blue and with DFT in green.}
	\label{fig-cost-fp-pinball}
\end{wrapfigure}

For the results reported in this work, we ran 2'503 SCF-calculations, 5'214 variable-cell relaxations, 171'370 molecular dynamics simulations in the pinball model, and 11'525 FPMD calculations, also 
\clearpage
\begin{wrapfigure}{r}{0.55\hsize}
	\includegraphics[width=\hsize]{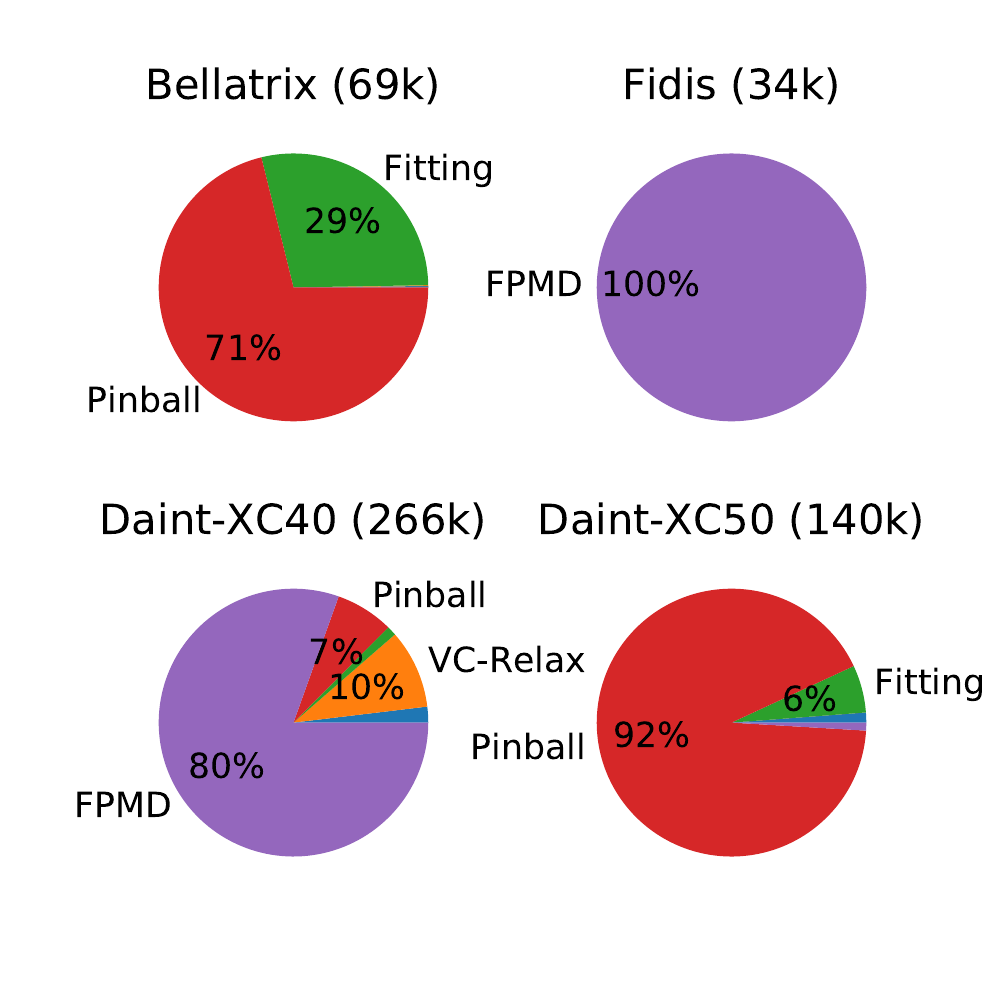}
	\caption{
	The total node hours used in each cluster is above every pie chart, which gives the relative usage for each type of calculation in this cluster.
	The pinball simulations are in red, all simulations that were done for the fitting of the model in green,
	variable-cell relaxations are shown in orange, and single point SCF-calculations in blue.}
	\label{fig-costs}
\end{wrapfigure}
\noindent
counting restarts.
The calculations were performed on four different clusters:
The Bellatrix cluster of EPFL, having computing nodes of two Intel\textsuperscript{\tiny\textregistered} Sandy Bridge processors running at 2.2~GHz, with eight cores each;
The Fidis cluster of EPFL, with two Intel\textsuperscript{\tiny\textregistered} Broadwell processors running at 2.6 GHz, with 14 cores each;
The XC40 partition of the Piz Daint cluster at the Swiss National Supercomputing Centre (CSCS), with compute nodes of two Intel\textsuperscript{\tiny\textregistered} Xeon E5-2695 v4 with 18 cores each, running at 2.1~GHz;
The XC50 partition of the Piz Daint cluster at CSCS, with nodes of 12 Intel\textsuperscript{\tiny\textregistered} Xeon E5-2690 v3 at 2.60~GHz processors.
We give the computational cost of the simulations in \myfigref{fig-costs}, discerning by computer and calculation type.
The dominant calculations are the pinball simulations and the FPMD.
In \myfigref{fig-cost-fp-pinball} we show a histogram of the average node-time per ionic step for all the structures that were successfully fitted, revealing that the computational cost of the pinball model is about four orders of magnitude lower.
The computational

\noindent
efficiency was utilized to get converged statistics for 
all structures that were simulated with the pinball model, resulting in a larger total simulation time, which explains why the total computational cost is en par with FPMD.
It is evident that screening the
same number of structures just with FPMD would not have been possible with today's computer performance.

\clearpage

\bibliographystyle{supershort}
\bibliography{screening}
\end{document}